\documentclass[10pt]{article}
\usepackage[]{psfig}
\newcommand{\be}{\begin{equation}}
\newcommand{\ee}{\end{equation}}
\newcommand{\ba}{\begin{eqnarray}}
\newcommand{\ea}{\end{eqnarray}}

\newcommand{\adg}{{\hat{a}^\dagger}}

\newcommand{\n}{\nonumber\\}

\newcommand{\ks}{\not \! k}
\newcommand{\cs}{\not \! c}
\newcommand{\ds}{\not\!\! D}

\begin{document}
%

%\catchline{}{}{}{}{}
%
%%%%%%%%%%%%%%%%%%%%%%%%%%%%%%%%%%%%%%%%%%%%%%%%%%%%%%%%%%%%%%%%%%%%

\title{THREE LECTURES ON\\NONCOMMUTATIVE FIELD THEORIES\footnote{Lectures given at
the II International Conference on Fundamental Interactions, Pedra
Azul, Brazil, June 2004.}}

\author{F.A.~SCHAPOSNIK\footnote{Associated to CICBA.}
\\
{\normalsize\it Departamento de F\'\i sica, Universidad Nacional
de La Plata}\\ {\normalsize\it C.C. 67, 1900 La Plata,
Argentina}\\
}
\date{\hfill}
\maketitle

%\maketitle

%\pub{Received (Day Month Year)}{Revised (Day Month Year)}

\begin{abstract}
Classical and quantum aspects of noncommutative field theories are discussed. In
particular noncommutative solitons and instantons are constructed and also $d=2,3$
noncommutative fermion and bosonic (Wess-Zumino-Witten and Chern-Simons)
theories are analyzed.
\end{abstract}

\section{Introduction}

The development of Noncommutative Quantum Field Theories has a
long story that starts with Heisenberg observation (in a letter he
wrote to Peierls in the late 1930\cite{P}) on the possibility of
introducing {\it uncertainty relations for coordinates}, as a way
to avoid singularities of the electron self-energy. Peierls
eventually made use of these ideas in work related to the Landau
level problem. Heisenberg also commented on this posibility to
Pauli who, in turn, involved Oppenheimer in the
 discussion\cite{O}. It was finally Hartland Snyder, an student of
Openheimer, who published the first paper on {\it Quantized Space
Time}\cite{S}. Almost immediately C.N.~Yang reacted to this paper
publishing a letter to the Editor of the Physical Review\cite{Y}
where he extended Snyder treatment to the case of curved space (in
particular de Sitter space). In 1948 Moyal addressed to the
problem using Wigner phase-space distribution functions and he
introduced what is now known as the Moyal star product, a
noncommutative associative product, in order to discuss the
mathematical structure of quantum mechanics\cite{M}.

The contemporary success of the renormalization program shadowed
these ideas for a while. Mathematicians, Connes and
collaborators in particular, made however important advances
in the 1980's,
in a field today known as noncommutative geometry\cite{C}. The
physical applications of these ideas were mainly centered in
problems related to the standard model until Connes, Douglas and
Schwartz observed that noncommutative geometry arises as a
possible scenario for certain low energy limits of string theory
and M-theory\cite{CDS}. Afterwards, Seiberg and Witten\cite{SW}
identified limits in which the entire string dynamics can be
described in terms of noncommutative (Moyal deformed) Yang-Mills
theory. Since then, more than 1500 papers (not including the
present one) appeared in the {\it arXiv} dealing with different
applications of noncommutative theories in physical problems.

Many of these recent developments, including Seiberg-Witten work,
were triggered in part by the construction of noncommutative
instantons\cite{NS} and solitons\cite{GMS}, solutions to the
classical equations of motion or BPS equations of noncommutative
 theories. One of the aims of these lectures is, precisely,
 to describe the construction of explicit vortex solutions for
the noncommutative version of the Abelian Higgs model and of
instanton solutions for noncommutative Yang-Mills theory. Appart
from these classical aspects of noncommutative field theories, we
shall also describe some very interesting features of two and
three dimensional noncommutative quantum field theories. In
particular, will discuss the noncommutative version of the
Wess-Zumino-Witten model and its connection with fermion models.
Also, we will analyze  three dimensional fermionic models and
their connection with the noncommutative version of Chern Simons
theory.

~

The plan of the three lectures presented in sections 2, 3 and 4 is the following:

~

\begin{itemize}
\item {{\bf Noncommutative Field Theories: tools} }

\noindent - Introduction \hfill 3

\noindent - Configuration space $\leftrightarrow$ Fock space
\hfill 5

\noindent - Noncommutative gauge theories \hfill 10

\noindent - Noncommutative field theories in curved space \hfill
14

\vspace{0.3 cm}

\item{{\bf Noncommutative solitons and instantons}
}

\noindent - Noncommutative vortices \hfill 17

\noindent - Noncommutative instantons\hfill 23

\noindent - Monopoles\hfill 29

\vspace{0.3 cm}

\item{{\bf Noncommutative Theories in d=2,3 dimensions}
}

\noindent - The Seiberg-Witten map \hfill 33

\noindent - Fermion models in two dimensional space  and the W*Z*W
model \hfill 34

\noindent - C*S theory in d=3 dimensions \hfill 43
\end{itemize}

~

\noindent Bibliography: There are many very good lectures and
review articles which cover the basic aspects discussed in Lecture
1. In particular those of .~A.~Harvey, {\it Komaba lectures on
noncommutative solitons and D-branes},  arXiv:hep-th/0102076;
M.~R.~Douglas and N.~A.~Nekrasov, {\it Noncommutative field
theory},  Rev.\ Mod.\ Phys.\  {\bf 73}, 977 (2001); R.~J.~Szabo,
{\it Quantum field theory on noncommutative spaces} Phys.\ Rept.\
{\bf 378}, 207 (2003). Concerning the specific results described
in Lectures 2 and 3, they were obtained in collaboration with
D.H.~Correa, P.~Forgacs, C.D.~Fosco,  N.~Grandi, G.S.~Lozano, E.F.~Moreno,
M.J.~Rodr\'\i guez, G.A.~Silva and  G.~Torroba and originally presented in
references [11]-[16].

\section{Noncommutative Field Theories: tools}

\subsection*{Introduction}

There are different settings for noncommutative field theories
(NCFT). The one that has been most used in all recent applications
 is based in the so-called Moyal product in which
for all calculation purposes (differentiation, integration, etc)
the space-time coordinates are treated as ordinary (commutative)
variables and noncommutativity enters into play  in the way
 in which   fields are multiplied. As an example,
 consider a typical (ordinary)
action defined on ${R}^n$  governing the dynamics of two scalars
\ba S &=& \int d^4x \left( \frac{1}{2} \partial_\mu \phi(x)
\partial^\mu \phi(x) \frac{1}{2} \partial_\mu \chi(x)
\partial^\mu \chi(x) + \frac{m^2}{2} \phi(x) \phi(x) +  \right.
\nonumber\\
&& g_3\phi(x) \phi(x)\chi(x) + \left. g_4 \phi(x) \phi(x)\chi(x) \chi(x) + \ldots\right)
\label{lc}
\ea
The noncommutative version of this action is built by replacing the
ordinary product among fields by a certain noncommutative product represented by
the  symbol $*$,
\ba
S_{nc} &=& \int d^4x \left( \frac{1}{2} \partial_\mu \phi(x) *\partial^\mu \phi(x)
\frac{1}{2} \partial_\mu \chi(x) * \partial^\mu \chi(x)
+ \frac{m^2}{2} \phi(x) * \phi(x) +  \right.
\nonumber\\
&&g_3\phi(x)* \phi(x)*\chi(x) +
 \left. g_4 \phi(x)* \phi(x)*\chi(x)* \chi(x) + \ldots\right)
\label{lnc}
\ea
What
has this modification to do with noncommutative geometry? To answer this question,
one can think the product
$\phi(x) *\chi(x)$ as the result of a deformation of the algebra of functions on
 $R^n$ to a noncommutative algebra ${\cal A}$. Such deformation can be
 connected to noncommutative geometry if one defines a Lie algebra for coordinates
 $x^\mu$ in  $R^n$ in the form
 \be
 [x^\mu,x^\nu] = i \theta^{\mu\nu}
 \label{unos}
 \ee
with ${\cal A}$ such that
\be
\phi(x) *\chi(x) = \phi(x) \chi(x) + \frac{i}{2}\theta^{\mu\nu}\partial_\mu \phi(x)
\partial_\nu \chi(x) + {\cal O}(\theta^2)
\label{pois}
\ee
Here $\theta^{\mu\nu}$ is a constant antisymmetric matrix of rank $2r
\leq d$ and dimensions of $(length)^2$. All terms in (\ref{pois}) are local
differential bilinears in $\phi$ and $\chi$. From (\ref{pois}) one sees that
the $*$ commutator
\be
[\phi,\chi] =\phi(x) *\chi(x)- \chi(x) *\phi(x) = i\theta^{\mu\nu}\partial_\mu \phi(x)
\partial_\nu \chi(x) + {\cal O}(\theta^2)
\label{pois2}
\ee
defines, up to order $\theta$, a Poisson structure. At order $\theta$ such a Poisson
structure satisfies Jacobi identity and hence associativity of the resulting
$*$ product. Demanding
associativity to all orders in $\theta$, one unavoidably
ends with the product defined by Moyal in 1949\cite{M}
\be
\phi(x)*\chi(x) = \phi (x) e^{\frac{i}{2} \theta^{\mu
\nu}\overleftarrow{\partial_\mu}
\overrightarrow{\partial_\nu }}\chi(x)
\label{moyalito}
\ee
which can be expanded as
\begin{eqnarray}
\phi(x)*\chi(x) = \phi(x)\chi (x) + \frac{i}{2} \theta_{\mu \nu}\partial^\mu
\phi (x) \partial^\nu\chi(x) - \frac{1}{8}
\theta_{\mu\nu}\theta_{\alpha
\beta}\partial_\mu\partial_\alpha\phi(x)
\partial_\nu \partial_\beta\chi(x) + \ldots \label{moyal}\nonumber\\
\end{eqnarray}

A very important property of the Moyal product is the following
\be
\phi(x)*\chi (x)= \phi(x)\chi (x) + \partial_\mu\Lambda_\mu(x)
\ee
with
\be
\Lambda_\mu = \frac{i}{2}
\theta_{\mu
\nu}\phi (x) \partial^\nu\chi(x) - \frac{1}{8}
\theta_{\mu\alpha}\theta_{\nu \beta}\partial_\nu\phi(x)
\partial_\alpha \partial_\beta\chi(x) + \ldots \nonumber\\
\ee
so that
\be
[\phi,\chi] = \partial_\mu\Sigma_\mu(x)
\ee
and hence if fields and their derivatives satisfy appropriate boundary
conditions one has
\be
\int d^4x \phi(x)*\chi (x) = \int d^4x \chi(x)*\phi (x) = \int d^4x \phi(x)*\chi (x)
\ee
Note that because of this property, quadratic terms in action (\ref{lnc}) coincide
with those in (\ref{lc}) and hence free commutative and noncommutative actions
(and hence propagators) are the same. It is through interactions that new features
should be expected.

One can also very easily verify that the $*$ product has the cyclic property
\be
 \int dx \,\phi(x)*\chi (x)*\psi(x) =
 \int dx\, \psi(x)*\chi(x) * \phi(x)
\ee
and that Leibnitz rule holds
\be
\partial_\mu \left(\phi(x)*\chi (x)\right) = \partial_\mu \phi(x)*\chi (x)
  + \phi(x)*\partial_\mu\chi (x)
\ee
We then see that for all purposes integration on $R^n$ corresponds
to a trace.

The $*$ product of two real fields is not necessarily real. However, the
definition of the  Moyal commutator
is consistent with the usual Hesienberg property for real
fields,
\be
\overline{[\phi,\chi]} = \overline{\phi*\chi} - \overline{\chi*\phi} = \bar \chi *
\bar\phi - \bar \phi*\bar \chi =  \chi *
\phi -  \phi*\chi = [\chi,\phi]
\ee

\subsection*{Configuration space $\leftrightarrow$ Fock space}
Let us specialize to two-dimensional space-time ($R^2$) with
coordinates $x^0$, $x^1$. In this case we can write \be
\theta^{\mu\nu} = \theta \varepsilon^{\mu\nu} \ee with $\theta$ a
constant and $\varepsilon^{01}= -\varepsilon^{10} = 1$.
Eq.(\ref{unos}) then reduces to \be [x^0,x^1] = \theta \ee or,
defining \be \hat a = \frac{x^0 + i x^1}{\sqrt{2\theta}} \; ,
\;\;\; \hat a^\dagger = \frac{x^0 - i x^1}{\sqrt{2\theta}} \ee
\be
[\hat a,\hat a^\dagger] = 1
\ee
Then $\hat a$ and $\hat
a^\dagger$ realize the algebra of  annihilation and creation
operators usually introduced in the process of second quantization.
One can then consider a Fock space with a basis $|n\rangle$
($n \geq 0$)
provided by the eigenfunctions of the number operator $N$,
\ba
\hat N &=& \adg \hat a
\label{seis}
\\
\hat N |n\rangle &=& n |n\rangle
\label{cincop}
\ea
with
\ba
\hat a |n\rangle &=& \sqrt{n} |n-1\rangle \; , \;\;\; n>0 \; ,\\
\adg |n\rangle &=& \sqrt{n+1}|n+1\rangle
\label{siete}
\ea
and the vacuum state $|0\rangle$ defined so that
\be
\hat a |0\rangle = 0
\label{ocho}
\ee

For $ \theta\ \simeq 0 $ one can write
\be
\hat N = {\hat a}^\dagger \hat a \approx  ( x^2 + y^2) /{2\theta}=
 {r^2}/{2\theta}\ee
so that  configuration
space at infinity  can be connected with   $n \to \infty$
in Fock space.

Let us now derive a   precise connection,
known as the Weyl connection, between the Moyal product of fields
in configuration space and the product of operators in
Fock space. To this end, consider
a field $f(x^0,x^1)$ in configuration space and take its Fourier
transform
\be
\tilde f(k,\bar k) = \int d^2z f(x^0,x^1) \exp
\left( i ( k_0 x^0 + k_1 x^1)
 \right)
 \ee
Then,  define an operator, acting in Fock space, associated to $\phi$,
\be
O_f(\hat a,\hat a^\dagger)  = \frac{1}{\theta}
\int \frac{d^2k}{(2\pi)^2}
\tilde f(k,\bar k)
\exp\left(-i\bar k \hat a + k \hat a^\dagger\right)
\label{trece}
\ee
At this point, one   can already  verify that
\be
\int d^2z f(z,\bar z) = 2\pi \theta {\rm Tr} O_f
\label{for}
\ee
where we have defined
\be
z = \frac{x^0 + i x^1}{\sqrt 2} \; , \;\;\;
\bar z = \frac{x^0 - i x^1}{\sqrt 2}
\ee
and ``Tr'' means the Fock space trace  of operator $O_f$.
To see this, we start from (\ref{trece})  and write
\ba
\langle n |O_f |n\rangle = \frac{1}{4\pi^2 \theta} \int \tilde
f(\bar k , k) \langle n|     \exp \left(-i(\bar k \hat a + k \adg)\right)
|n\rangle  \nonumber\\
\label{xx}
\ea
which, using the Baker-Campbell-Haussdorff    formula, can be arranged as
\be
O_{nn} =   \frac{1}{4\pi^2 \theta} \int \tilde
f(\bar k , k)\exp \frac{\bar kk}{2}
\langle n | \exp(-ik \adg ) \exp(-i\bar k \hat a) |n\rangle
\ee
Then, using the Schwinger formula\cite{Schw}
%R
\be
\langle n | \exp(-ik \adg ) \exp(-i\bar k \hat a) |n\rangle
= L_n(\bar k k)
\ee
with $L_n$ the Laguerre polynomial one gets
\be
{\rm Tr} O_f = \sum_n O_{nn} = \frac{1}{4\pi^2 \theta}
\int d^2 k \tilde f(\bar k,k)\exp \frac{\bar kk}{2} \sum_n
L_n(\bar k k)
\ee
Using the identity
\be
\sum_n t^n L_n(x^)   = \frac{1}{1-t} \exp \left(
\frac{tx}{1-t}
\right)
\ee
one gets
\be
\sum_n L_n(\bar k k) = 2 \pi \delta(k) \delta(\bar k)
\ee
this leading to formula (\ref{for})
\be
{\rm Tr} O_f =  = \frac{1}{2\pi \theta} \tilde f(0,0) =
\frac{1}{2\pi \theta} \int d^2 z f(\bar z,z)
\ee

The natural   basis to use in Fock space in order to expand
operators $O$
consists of the elementary operators $|k\rangle \langle l|$,
\be
O_f = \sum_{kl} \left(O_f\right)_{kl}|k\rangle \langle l|
\ee
The basis operators can be in turn expressed in terms of $\adg$
and $\hat a$ in the form
\be
|k\rangle \langle l| =
 : \frac{{\hat a}^{\dagger k}}{\sqrt {k!}}
\exp(-\adg \hat a)
\frac{{\hat a}^l}{\sqrt {l!}}:
\label{pro}
\ee
where $:~:$ denotes normal ordering. That identity (\ref{pro})
holds can be seen just by verifying that, when acting on kets
$|p\rangle$'s and on bras $\langle q |$'s, both sides give the same
answer.

Expression (\ref{trece}) gives a symmetric ordered operator. We
can write an analogous formula but for  a \underline{normal ordered}
operator just by using the Baker-Cambell-Hausdorff relation. One
has, starting from (\ref{trece})
\ba
:O_f(\hat a, \adg)\!: &=&
\frac{1}{4\pi^2 \theta} \int d^2k \tilde
f_N(k,\bar k) :\exp \left(-i(\bar k \hat a + k \adg)\right)\!:
\nonumber\\&=&
 \frac{1}{4\pi^2 \theta} \int d^2k \tilde f(k,\bar k)
 \exp \left(-i(\bar k \hat a + k \adg)\right) \exp\left(
\frac{k^2}{4}
\right )
\label{trece1}
\ea
Note that we use the subscript $N$ associated to the
normal-ordered expression.

Consider the operator $O_n = |n\rangle\langle n|$. For simplicity, we temporarily
put $\theta =1$ Using
representation (\ref{pro}), we have
\be
O_n = : \frac{{\hat a}^{\dagger n}}{\sqrt {n!}}
\exp(-\adg \hat a)
\frac{{\hat a}^n}{\sqrt {n!}}: = \frac{1}{4\pi^2} \int d^2k \tilde
g^n_N(k,\bar k) :\exp \left(-i(\bar k \hat a + k \adg)\right)\!:
\label{asi}
\ee
with
\be
\tilde g_N^n(\bar k, k) = \frac{1}{n!}\int d^2z
\exp\left(i (k \bar z + \bar k z)\right)
\bar z^n \exp(-\bar z z) z^n
\ee
We can use at this point an integral representation for the
Laguerre polynomials (see for example formula (B11) in the
very useful book on coherent states by A.~Perelemov\cite{Pere}),
\be
L_n\left(\frac{k^2}{2}\right) = \frac{1}{2\pi n!}
\exp\left(\frac{ k^2}{2}\right) \int d^2z |z|^{2n} \exp\left(i (k \bar z + \bar k
z)\right)
\ee
and use the second line in eq.(\ref{trece1}) to write
\be
|n\rangle \langle n| =     \frac{1}{2\pi} \int d^2 k
\exp\left(-\frac{ k^2}{4}\right)L_n\left(\frac{k^2}{2}\right)
\exp \left(-i(\bar k \hat a + k \adg)\right) \label{cuaren}
\ee

The function $g^n(x)$ that corresponds to the operator $O_n =
|n\rangle \langle n|$ can be copied from (\ref{cuaren})
\ba
g^n(x) &=& \frac{1}{2\pi} \int d^2k
\exp\left(-\frac{ k^2}{4}\right)L_n\left(\frac{k^2}{2}\right)
\exp(-ik.x) \nonumber\\ &=& 2(-1)^n \exp(-r^2) L_n(2r^2)
\label{casi}
\ea
where $r^2 = x^2 + y^2$. Re-introducing $\theta$  we then have
the connection
\be
|n \rangle \langle n| \to     2(-1)^n \exp\left(- \frac{r^2}{\theta}\right)
L_n\left(\frac{2r^2}{\theta}\right)
\label{sinn}
\ee
Now, we are ready to present the most significant formula in this section,
\be
\underbrace{O_f O_g}_{operator~product} =
O_{\!\!{\underbrace{f*g}_{*\,product}}}
\label{conne}
\ee
It  shows that the  star product of fields in configuration space as
defined in (\ref{moyalito})
becomes   a simple operator product in Fock space. In this way, one can either work
using Moyal products or operator products and pass from one language to the other
just by Weyl (anti)transforming the results.

In order to  prove (\ref{conne}) we start from the l.h.s and use
(\ref{trece1}) to write
\ba
O_f \cdot O_g &=&    \!\!\!
\frac{1}{16\pi^4 \theta} \int d^2k\int d^2k' \tilde f(k,\bar k)  \tilde g(k',\bar k')
\exp \left(-i(\bar k \hat a + k \adg)\right) \exp\left( \frac{k^2}{4}\right )
\nonumber\\
&&\exp \left(-i(\bar k' \hat a + k' \adg)\right) \exp\left( \frac{{k'}^2}{4}\right )
\nonumber\\
&=&\!\!\!  \frac{1}{16\pi^4 \theta} \int d^2k\int d^2k'
\tilde f(k,\bar k) \tilde g(k',\bar k')
\exp \left(-i\left((\bar k + \bar {k'}) \hat a + (k+k')  \adg\right)\right)
\nonumber\\
&&
  \exp\left( \frac{{k}^2}{4}\right )\exp\left( \frac{{k'}^2}{4}\right )
  \exp\left(\frac{i}{2} \left(k \bar{k'} - k' \bar k \right)
  \right)
  \label{porq}
 \ea
We now proceed to the change of variables  $k + k' = p$ , $(k-k')/2
= q$ and analogously for "bar" variables. Then  (\ref{porq})
becomes
\ba
O_f \cdot O_g &=&    \!\!\!
\frac{1}{4\pi^2}\int d^2p   \exp \left(-i(\bar p \hat a + p \adg)\right)
\left[
\frac{1}{4\pi^2}\int d^2q   \exp\left(\frac{i}{2} \left(q \bar{p} - p \bar q \right)
\right) \right.
\nonumber\\
&&
\left.\vphantom  {\frac{1}{4\pi^2}\int d^2q}
 \tilde f(q + p/2,\bar q + \bar p/2) \tilde g(-q + p/2,-\bar q + \bar p/2)
\right]
  \ea
Now, the factor in square brackets is nothing but the Fourier
transform of $f*g$ for noncommutative ${R}^2$ (and with $\theta =
1$). Hence, we end with
\be
 O_f \cdot O_g =
 \frac{1}{4\pi^2}\int d^2p   \exp \left(-i(\bar p \hat a + p \adg)\right)
\widetilde{f*g}(p,\bar p) = O_{f*g}
 \ee
So, we have established the announced  connection between operator
multiplication and the star product of functions.

It is easy to identify the operation that
corresponds to differentiation  in in Fock space. Indeed, starting from
\be
[\hat a^\dagger, \hat a^n] = -n \hat a^{n-1}
\label{unosd}
\ee
we see that for any ``holomorphic'' function $f(a)$ written in the   form
\be
f(\hat a) = \sum c_n \hat a^n
\ee
one can define a differentiation operation through the formula
\be
\frac{\partial f}{\partial a} =  - [\hat a^\dagger,f(\hat a)]
\ee
and analogously for any $f(\hat a^\dagger)$. Then, differentiation of a field
$\phi(z,\bar z)$ becomes, in operator language,
\be
\partial_z \phi(z,\bar z)  \to  -\frac{1}{\sqrt \theta}
[ \hat a^\dagger,O_\phi]\, , \;\;\;\;\;\; \;\;\;\;\;\;
\partial_{\bar z}\phi(z,\bar z) \to
 {\frac{1}{\sqrt \theta} [ \hat a,O_\phi]}
 \ee
and then  action (\ref{lnc}) can be written in operator language
 in the form
\ba
S_{nc}  &= & \pi \theta {\rm Tr}
 \left( [\hat a,O_\phi]^2 + [\hat a^\dagger,O_\phi]^2
 ([\hat a,O_\chi]^2 + [\hat a^\dagger,O_\chi]^2
\right.  \nonumber\\
&&  \left.{m^2} O_\phi^2 + {m^2} O_\chi^2
 +
2g_3O_\phi(x) O_\phi(x) O_\chi(x) +
2 g_4 O_\phi(x) O_\phi(x)O_\chi(x) O_\chi(x)\right)
 \nonumber\\
\label{lncop}
\ea
From here on we shall abandon the notation $O_\phi$ for operators
and just write $\phi$ both in configuration and Fock space. Also, hats in
operators will be discarded.

\subsection*{\it An example: the extrema of a symmetry breaking potential}

We are ready to  understand the difficulties and richness one encounters
when searching for noncommutative solitons just by  disregarding
kinetic energy terms   and studying the minima of
a typical symmetry breaking   potential,

\be V[\phi *  \phi] =  \frac{1}{2} m^2 \phi * \phi -
\frac{\lambda}{4} \phi*\phi*\phi*\phi \label{pot}
\ee
The   equation for  extrema  is
\begin{equation}
m^2 \phi - \lambda \phi*\phi*\phi = 0
\end{equation}
or, with the shift
$
\phi \to ({m}/{\sqrt \lambda}) \phi
$,
\begin{equation}
\phi(x)*\phi(x)*\phi(x) = \phi(x)  \label{chiste}
\end{equation}
One can find a subset of solutions
 if one considers, instead of (\ref{chiste}),
 the quadratic equation
\begin{equation}
 \phi_0(x) * \phi_0(x) =  \phi_0(x)
 \label{2}
\end{equation}
since, evidently,  $\phi = \phi_0(x)$ satisfies
(\ref{chiste}).
Although simpler than the cubic equation (\ref{chiste}),
 eq.(\ref{2}) also implies, through Moyal star products,  derivatives
 of all orders and then only a few
  solutions can be readily found  with some little work.
For example, in $d=2$ dimensions one finds
\begin{eqnarray}
\phi_0 &=& 0 \nonumber\\
\phi_0 &=& 1 \nonumber\\
\phi_0 &=& \frac{2}{\sqrt\theta}
{\exp\left(-\vec x^2/\theta
\right)}_{~ ~ ~ \overrightarrow{\theta \to
0} ~ ~ ~} \delta^{(2)}(x) \label{primera}
\end{eqnarray}
Already the exponential solution in (\ref{primera}) shows that nontrivial regular
solutions, which
were excluded for such a model
in the commutative space due to Derrick theorem, can
be found in noncommutative space. The reason for this is clear:
the presence of the  parameter $\theta$ carrying
 dimensions of $length^2$,   prevents the  Derrick scaling
 analysis leading to the
negative results in ordinary space.

To look for more general solutions, let us restrict the search to
$ {R}^2$ and note that eq.(\ref{2}) looks like a projector
equation in configuration space. Now, in Fock space, it is very
easy to write projectors,
\be
P_n = |n\rangle\langle n| \; , \;\;\; P_n^2 = 1
\ee
which of course satisfy the analogous of eq.(\ref{chiste}),
\be
P_n^3 = P_n
\ee
So, we can say that we already found  a
solution to (\ref{chiste}) in operator form,
\be
O_\phi = |n\rangle \langle n|
\ee
or, through the Weyl connection (see (\ref{sinn})), in configuration space,
\be
\phi_0^n(\bar z,z) =
2 (-1)^n \exp\left( -\frac{\bar z z}{\theta}\right)
L_n\left(\frac{2\bar z z}{\theta}\right)
\label{ficero}
\ee

We then conclude that
in $d=2$ dimensions, a family of solutions of eq.(\ref{chiste}) is given by

\begin{eqnarray}
P_\phi &=& \sum \lambda_n |n\rangle \langle n| \hspace{1cm}{\rm in~Fock~space}
\nonumber\\
\phi & =
&\sum \lambda_n \phi_0^n(\vec x)
\hspace{1.015cm} {\rm in~configuration~space}\nonumber\\
\label{sumas}
\end{eqnarray}
with $ \lambda_n = 0, \pm 1 $ and $\phi_0^n$ given by (\ref{ficero}).

\subsection*{Noncommutative gauge theories}

Let us start by considering the noncommutative version of a pure
$U(1)$ gauge theory. Given the gauge connection $A_\mu(x)$, we
have first to define an infinitesimal gauge transformation with
parameter $\epsilon(x)$. It is more or less evident that the naive
transformation law
\be
\delta A_\mu = \partial_\mu \epsilon(x)
\label{naive}
\ee
will not work in noncommutative
space. Instead, and inspired in non-Abelian gauge theories
in ordinary space, one defines
\be
\delta A_\mu = \partial_\mu \epsilon(x) +i\left(A_\mu*\epsilon - \epsilon*A_\mu
\right)
= \partial_\mu \epsilon(x) +i[A_\mu,\epsilon]
\label{naivn}
\ee
which corresponds, for finite gauge transformations to a
transformation law of the form
\be
A_\mu^g = g^{\dagger} * A_\mu * g + \frac{1}{i} g^{\dagger}* \partial_\mu g
\ee
with
\be
g(x) = \exp_*(i\epsilon(x)) = 1 + i \epsilon(x) - \frac{1}{2}
\epsilon(x)*\epsilon(x) + \ldots
\ee
We see that elements of the $U(1)$ gauge group are now $*$-exponentials and to
emphasize this fact
we shall write $U_*(1)$ for the gauge group. Of course one has to define
the inverse of an element $g$ through $*$ product equations,
\be
g(x)*g^{-1}(x) = g^{-1}(x)*g(x) = 1
\ee
and one has $g^{-1} = g^\dagger$.

The covariant derivative inferred  from (\ref{naivn}),
\be
D_\mu[A] =  \partial_\mu  +i[A_\mu,~]
\ee
leads to
\be
[D_\mu,D_\nu]f = i [F_{\mu\nu},f]
\ee
with  the curvature $F_{\mu\nu}$ given by
\be
F_{\mu\nu} = \partial_\mu A_\nu - \partial_\nu A_\mu + i
[A_\mu,A_\nu]
\ee
Such $U_*(1)$ field strength is not gauge invariant but
transforms covariantly,
\be
F_{\mu\nu}^g = g^{-1} * F_{\mu \nu} * g
\ee
Of course the noncommutative version of the Maxwell action,
\be
S=-\frac{1}{4} \int d^dx F_{\mu\nu}* F^{\mu \nu} =
-\frac{1}{4}\int d^dx F_{\mu\nu} F^{\mu \nu}
\ee
is  gauge-invariant since, as explained above, integration acts as a trace for
the $*$ product.

Gauge invariant couplings to matter fields can be  easily defined. One should observe,
however, that
even in the $U_*(1)$ case, one has the possibility
of  considering fields   in the ``fundamental
representation'' ($f$),
defined through the gauge transformation law
\be
\psi \to \phi^g = g * \phi
\label{unat}
\ee
or in the ``anti-fundamental representation'' ($\bar f$)
\be
\chi \to \chi^g =  \chi* g^{-1}
\label{dost}
\ee
One can even define an ``adjoint $U_*(1)$ representation'' ($ad$) in which
fields transform as
\be
\eta \to \eta^g = g * \eta * g^{-1}
\label{trestt}
\ee

Consider for example Dirac fermion fields coupled to a gauge field. A gauge invariant
Dirac action can be written in the form

\[
S[\bar \psi,\psi,A] = \int d^dx  \bar \psi * i \gamma^\mu D_\mu
\psi \]
with the covariant derivative   chosen according to the
fermion representation,
\begin{eqnarray}
D^f_\mu[A]\psi(x) &=&
 \partial_\mu \psi(x)  + i A_\mu(x)*\psi(x)\label{d1}\\
  D^{\bar f}_\mu[A]\psi(x) &=&
\partial_\mu \psi(x) - i \psi(x) *A_\mu(x) \label{d2}\\
D^{ad}_\mu[A]\psi(x) &=&  \partial_\mu \psi(x)  + i[ A_\mu(x),\psi(x)]
\label{d3}
\end{eqnarray}
In many ways the noncommutative $U_*(1)$ theory behaves like a
non-Abelian gauge theory in ordinary space, one with structure
constants depending on the momenta. In particular, the fact that
in non-Abelian theories  the charge of matter fields is fixed by
the representation of the fields, here corresponds to the fact
that fermions carry only $+1$ charge in the fundamental, $-1$ in the
anti-fundamental and $0$-charge in the adjoint (although in this
last case carries a ``dipole'' moment).

Let us now extend this results to the case of a non-Abelian gauge
symmetry. For this, one has to take into account
that the $*$ product will in general destroy the closure condition and
hence not all the Lie groups can be chosen as gauge groups. As an
example, consider  traceless hermitian $x$-dependent $n\times n$
matrices $A(x)$ and $B(x)$, elements  of ${\cal SU(N)}$,
 the Lie
algebra of $SU(N)$. One can easily check that $A*B - B*A$ is not
traceless anymore. This poses a problem when on considers a non Abelian field strength
which includes a commutator of gauge fields.
 Indeed, one has, in the $SU(N)$ case,
\be
[A_\mu,A_\nu] = A_\mu^a * A_\nu^b t^a t^b - A_\nu^b * A_\mu^a  t^b t^a
\label{conmu}
\ee
Now, the $\cal SU(N)$ generators satisfy
\be
t^a t^b = 2i f_{abc} t^c + \frac{2}{N} \delta_{ab}
{I} + 2 d_{abc}t^c
\ee
with $d_{abc}$ the completely symmetric $SU(N)$ tensor. Then, commutator (\ref{conmu})
and, {\it a fortiori}, the
curvature, does not belong to the ${\cal SU(N)}$ algebra since it has, in principle, a
component along the identity,
\be F_{\mu\nu} = ~^{(1)\,}\!\!F_{\mu\nu}^a t^a +
^{(2)}\!\!F_{\mu\nu} I
 \ee
 One can avoid this problem
by considering from the start the $U_*(N)$ gauge group which includes
one more generator identified with the identity.
There are other proposals to evade  this problem by
considering other gauge groups. For example, one
can pose some constraints on gauge potentials and
gauge transformations so that
when the deformation parameter $\theta$ vanishes one
recovers the ordinary orthogonal and
symplectic gauge theories\cite{og}. We shall not discuss these issues here.

Another peculiarity of noncommutative gauge theories is that one
cannot straightforwardly define local gauge invariant observables
by taking the trace of gauge covariant objects. The cyclic
property of the group trace can not be used when fields are
multiplied via the Moyal product except when an integral over
space-time allows to simultaneously exploit cyclicity of the $*$-product. That
is, one can easily write gauge invariant objects like
\begin{eqnarray}
{\cal O} &=& {\rm tr}
\int d^d x \left( F_{\mu\nu} * F_{\mu\nu}\right)^n
\to {\rm tr} \int d^dx g^{\dagger} *  \left( F_{\mu\nu} *
F_{\mu\nu}\right)^n * g
\nonumber\\ &=&  {\rm tr}\int d^d x
\left( F_{\mu\nu} * F_{\mu\nu}\right)^n
\label{obs}
\end{eqnarray}
but defining local gauige-invariant objects pose problems. Even
when
constructing Wilson loop operators one faces difficulties. Consider
a path $L$, starting at $x_1$ and ending at $x_2$,
one can define the holonomy $W_L$  as in ordinary gauge
theories,
\be
W_{L(x_1,x_2)} = P \exp_*\left(i\int_L dx^\mu A_\mu
\right)
\ee
where $\exp_*$ indicates that products in the expansion of the path-ordered exponential
being $*$-products. Under gauge transformations, $W_L$ transforms
according to
\be
W_{L(x_1,x_2)} \to g^{\dagger}(x_1) * W_{L(x_1,x_2)} * g(x_2)
\label{have}
\ee
If the path is closed one has $x_1 = x_2$ but again one faces the
problem that just taking the group trace of $W_L$ does not define
a gauge invariant operator as it does in ordinary gauge theories
since one cannot cyclically permute $g$ elements to cancel them.

One proposal to overcome such problems consists in defining open
Wilson lines. More precisely, one parametrizes the path L(x, x + v) by
smooth embedding functions $\xi^\mu(t)$ with $t \in [0,1]$ such
that $\xi^\mu(0) = 0$ and $\xi^\mu(1) = v^\mu$. The holonomy over
such a contour is then
\begin{eqnarray}
W_{L(x,x+v)}&=& P \exp_* \left(i \int_L d\xi^\mu A_\mu(x +
v)\right)\nonumber\\
&=& 1 + \sum_{n=1}^\infty i^n \int_0^1 dt_1\int_{t_1}^1 dt_2
\ldots \int_{t_{n-1}}^1 dt_{n} {\dot \xi}^{\mu_1}(t_1) \ldots
{\dot \xi}^{\mu_n}(t_n) \nonumber\\
&\times& A_{\mu_1}(x + \xi(t_1))* \ldots  *A_{\mu_n}(x + \xi(t_n))
\end{eqnarray}
Under a gauge transformation,  one has, according to (\ref{have}),
\be
W_{L(x,x+v)} \to g^{\dagger}(x) * W_{L(x,x+v)} * g(x+v)
\label{have2}
\ee
Now, a gauge invariant observable can be constructed from $W_L$ in
the form
\be
{\cal O}(L(x,x+v)) = \int d^dx \,{\rm tr} \left( W_{L(x,x+v)}\right)*
\exp\left(i k_\mu(v)x^\mu\right)
\ee
where
\be
k^\mu(v) = \left( \theta^{-1}\right)_{\mu\nu} v^\nu
\ee
Gauge invariance of ${\cal O}(L(x,x+v)$ is ensured because of the
fact that
\be
\exp\left(i k_\mu(v)x^\mu\right) * g(x) * \exp\left(-i k_\mu(v)x^\mu\right)
= g(x +v)
\ee
together with cyclicity of the trace and the integrated $*$
product.
Strikingly enough, one can construct for non-commutative gauge
theories gauge invariant observables associated with
{\it open} contours, in contrast with ordinary gauge
theories where only closed contours are allowed.

I shall end this discussion of noncommutative gauge theories by discussing
  a very striking property of certain gauge field configurations that
I shall illustrate here with a two dimensional example. Consider a linear potential
in the Lorentz gauge,
\be
A_i = -\frac{B}{2} \varepsilon_{ij} x^j
\label{a2}
\ee
The corresponding field strength reads
\be
F_{12} = \partial_1A_2 - \partial_2A_1 - (A_1*A_2 - A_2*A_1) = B\left(1 +
\frac{B^2 \theta}{4}\right)
\ee
The first term in the parenthesis is  the constant result one
obtains for the field strength in ordinary space.
Noncommutativity shifts this value to a $\theta$-dependent one. Consider now
a gauge transformation of (\ref{a2}),
\be
A^g_i = g * A_i * g^{-1} - i\left( \partial_i g\right) * g^{-1}
\ee
which can be   rewritten as
\be
A^g_i = A_i + \left([g, A_i]  - i\left( \partial_i g\right)
\right) * g^{-1}
\ee
From (\ref{a2}) and the commutation relation
\be
[x^1,x^2] = i\theta
\ee
one has
\be
[g, A_i] =\frac{B}{2} \varepsilon_{ij}[x_i, g] = \frac{B}{2}\varepsilon_{ij}
 i \frac{\theta}{2} \varepsilon_{pq} \partial_p x_j \partial_q g = -\frac{i}{2}
 B\theta \partial_i g
 \ee
 and hence
 \be
 A^g_i =  = A_i -i\left(1 + \frac{\beta \theta}{2}\right)\partial_ig
 \label{parcero}
 \ee
 Then, for $B\theta = -2$ one has $A^g_i   = A_i$ so that
 the orbit consist of just one point! We shall see that this kind of configurations are
 relevant in the study of vortex and monopole solutions.

\subsection*{ Noncommutative field theories in curved space}
The problem of noncommutative field theories in a nontrivial background metric
needs some attention since, as it is evident, it implies a non-constant deformation
matrix $\theta_{\mu\nu} = \theta_{\mu\nu}(x)$. But once the dependence on $x^\mu$
of the deformation
parameter is admited, a closed, explicit Moyal-type formula is no more available and
in fact, defining a noncommutative associative product becomes rather involved.
Kontsevich\cite{kon} has been able to find the conditions under which
this is possible and has given a recipe to multiply
fields, order by order in $\theta$, which can not be written in a closed
form, as in the Moyal case. We shall not discuss this approach here but consider
a case in which important simplifications take place\cite{FCST}.

Consider a Poisson structure defined in the form
\be
\{\phi,\chi\} = \theta_{\mu\nu} \partial_\mu \phi \partial_\nu \chi
\ee
We then introduce a noncommutative product (which we denote with $\star$
to distinguish it from the constant $\theta$ Moyal $*$ product)
which, up to order $\theta$ takes
the form
\be
f \star g \equiv fg + \frac{i}{2}\{f,g\} + {\cal O}(\theta^2)
\label{starr}
\ee
One can see that the Jacobi identity for the Poisson bracket will ensure associativity of
the $\star$ product (\ref{starr})\cite{Cor2,FCST}. This implies that
\be
\theta^{ij}\partial_j\theta^{kl} +
\theta^{lj}\partial_j\theta^{ik} +
\theta^{kj}\partial_j\theta^{li} = 0
\ee
which can be written covariantly in the form
\be
\theta^{ij}\nabla_j\theta^{kl} +
\theta^{lj}\nabla_j\theta^{ik} +
\theta^{kj}\nabla_j\theta^{li} = 0
\ee
This equation is evidently satisfied whenever the following condition holds,
\be \nabla_i \theta^{jk}(x) = 0
 \label{suff}
\ee
This condition is very simple to handle  in two
dimensional space. Indeed, the most general $\theta^{ij}$ can be written
 in $d=2$ in the form
\be
 \theta^{ij} ( x^1, x^2) = \frac{\varepsilon^{ij}}{\sqrt {g(x)}}
\theta_0(  x^1,  x^2)
\label{clase1}
\ee
where $\theta( x^1, x^2)$ is a scalar. But
then, condition (\ref{suff}) reduces to
\be
\nabla_i  \left(\frac{\varepsilon^{jk}}{\sqrt{g(x)} }
\theta(x^1, x^2) \right)= \frac{\varepsilon^{jk}}{\sqrt{g(x)}}
\nabla_i
\theta (x^1,x^2) =
 \frac{\varepsilon^{jk}}{\sqrt {g(x)} }\partial_i
  \theta_0 (x^1,x^2) = 0
\label{conchap}
\ee
and hence $\theta_0$ must be a constant which one can normalize
(in some appropriate units) to
1.
Then,
 the sufficient condition
(\ref{suff}) is equivalent, in $d=2$ dimensions, to
\be \theta^{ij}(x^1,x^2) =
\frac{\varepsilon^{ij}}{\sqrt {g(x)}}
\ee
Let us  note at this point that in the presence of a metric, the
natural integration measure should be
\be
 d\mu(x) =  d^2 x \sqrt g
\ee
and this, as we shall see, is consistent with the definition of
a noncommutative product in which integrals act as a trace. Additional
results on this issue were obtained after these lectures were
presented\cite{FCST}. I briefly discuss below those which are relevant to the present
talk.

\subsection*{{\it The simple case $\theta = \theta(x^1)$}}

Let us specialize to the case in which $\theta$ depends just on one coordinate,
say $x^1$
(or, what is the same, the determinant of the background metric is
$g = g(x^1)$),

 \be [ x^1, x^2]_{ \star} = x^1 \star x^2  - x^2
\star x^1  = i \theta(x^1)
\label{11}
\ee
where $\star$ is the noncommutative associative
product   defined, up to order $\theta$  by eq.(\ref{starr})
Inspired in the change of variable used to study vortices in
curved space\cite{Cor2}, we multiply (\ref{11})
from the left and from the right  with
$1/\sqrt{\theta(x^1)}$,
\be
x^1 \star \frac{1}{\sqrt{\theta(x^1)}} \star x^2 \star
 \frac{1}{\sqrt{\theta(x^1)}}-
 \frac{1}{\sqrt{\theta(x^1)}}
\star x^2 \star  \frac{1}{\sqrt{\theta(x^1)}}
\star x^1 = i \label{2x}
\ee
so that if we change variables according to
\ba x &=& x^1
\nonumber\\
 y &=&  \frac{1}{\sqrt{\theta(x^1)}} \star x^2
\star  \frac{1}{\sqrt{\theta(x^1)}}\label{trescu} \ea
we have \be x \star  y - y \star x = i \label{cuatro} \ee
Note that variables $x$ and $ y$ as defined in (\ref{trescu}) are
hermitian.  Now, because of
 (\ref{cuatro}), the noncommutative $star$ product  can be
 realized as an ordinary Moyal product
$*$ with
 (constant) $\theta = 1$ and hence, in terms of variables $(x,y)$,
  we can proceed as one usually does.
In order to exploit the Moyal formula in terms of the original variables, let us
note that from (\ref{tres}) one has

\ba
x^1 &=& x \nonumber\\
x^2 &=& \sqrt{\theta( x)} * y * \sqrt{\theta( x)} = y\, \theta(x)
\ea
Then
the noncommutative product of two functions of the original variables
$x^1,x^2$ can be defined in the form
\ba  f( x^1, x^2) \star g(x^1,x^2) &\equiv&
f(  x,  y) *
g(  x,  y)
\nonumber\\
  &=&
f\left(  x, \theta(  x)   y
 \right)\exp\left(
\frac{i}{2} \frac{\overleftarrow{\partial}}{\partial   x}
\frac{\overrightarrow{\partial}}{\partial   y} - \frac{i}{2}
\frac{\overleftarrow{\partial}}{\partial   y}
\frac{\overrightarrow{\partial}}{\partial   x} \right)  \
  g\left(  x,\theta(  x)   y \right)\nonumber\\
 \label{cincos}
\ea
Now, we are ready to write integrals of product of fields provided
we work in terms of the new variables. For example one can write
\be I_2 = \int dx dy f(x,y)*g(x,y) \ee We can however rewrite
$I_2$ in term of variables $x^1,x^2$, using that \be
\frac{\partial y}{\partial x^2} = \frac{1}{\theta(x)} \ee so that
\be I_2 = \int dx^1dx^2 \frac{1}{\theta(x)} f(x^1,x^2)\star
g(x^1,x^2) \ee
so that we see that also in this approach, the natural integration
measure, in terms of variables $(x^1,x^2)$ is that given in (33),
\be d\mu(x) = dx^1 dx^2  \theta^{-1}(x) \ee

\section{Noncommutative solitons and instantons}

\subsection*{Noncommutative vortices}
Let us briefly review how   vortex solutions
were found in the Abelian Higgs model in ordinary
space\cite{NO,Bogo}. The energy
for
static, z-independent configurations is,
for the commutative version of the theory,
\be
E = \frac{1}{2} F_{ij}^2 + \overline{D_i\phi}{D_i\phi} +
\frac{\lambda}{4} (|\phi|^2 - \eta^2)^2
\label{energy}
\ee
Here $i=1,2$ and since fields depend just on two variables,
 one can consider the model in  two
dimensional Euclidean
space with
\be
D_i \phi = \partial_i - i A_i \phi \; , \;\;\; \phi = \phi^1 + i \phi^2
\ee
The Nielsen-Olesen strategy
to construct topologically regular solutions to the
equations of motion with finite energy (per unit length)
starts from a trivial (constant) solution
and implies the following steps:

~

\begin{enumerate}
\item{\underline{Trivial}}  solution
$
|\phi |= {\eta}  \; , \;\;\; A_i = {0}
$

~

\item {Topologically \underline{non-trivial} but  \underline{singular}
 solution}  (fluxon with $N$ units of magnetic flux) which in polar coordinates
 reads
\[
\phi =  {\eta} \exp(i N\varphi) \; , \;\;\; A_i =
n \partial_i \varphi\; , \;\;\;
\varepsilon_{ij}F_{ij} = 2\pi {N\delta^{(2)}(\vec x)}
\]

~

\item{ \underline{Regular}  Nielsen-Olesen vortex solution}
\be
\phi =  { f(r)}\exp(i N\varphi) \; , \;\;\; A_i =
 {a(r)} \partial_i \varphi
\ee
with
$
f(0)=a(0)= 0 \; , \;\;\; f(\infty) = \eta \; , a(\infty) = N
$
\end{enumerate}

\noindent  For the particular value $ \lambda = \lambda_{BPS} = 2 $
one can establish the Bogomol'nyi bound
$E \geq 2\pi N$,
attained when  first order
``Bogo\-mol'\-nyi''
equations hold
\begin{eqnarray}
F_{z \bar z} = \eta^2 - \bar \phi  \phi &~ ~ ~ &
-F_{z \bar z} = \eta^2 - \bar \phi \phi \nonumber\\
D_{\bar z} \phi = 0 ~ ~ ~ ~
 &~ ~ ~ & ~ ~ ~ ~ D_z \phi = 0 \nonumber\\
{\rm Self dual ~ ~ ~ ~ } &~ ~ ~ &
{\rm ~ ~ Anti selfdual}
\end{eqnarray}

One
can copy  this strategy, to attack the noncommutative
problem. The energy for the noncommutative version of the model
is obtained replacing ordinary products by $*$ products in eq.(\ref{energy}).
Since one is effectively working in $d=2$ dimensional
space,
one can attack the problem in Fock space and, at the end, rewrite the results
in configuration space.
Taking the scalars in the fundamental representation (but the other possibilities
go the same) one gets
\be
E = 2\pi \theta {\rm Tr} \left(
\frac{1}{2} B^2 + D_z \bar \phi D_{\bar z} \phi + D_z\phi D_{\bar z} \bar \phi
+ \frac{e^2}{2} \left(\phi \bar \phi - \eta^2
\right)
\right)
\label{lag}
\ee
where we have defined $B = iF_{z\bar z}$. For simplicity,
we have chosen the coupling constants
at the (ordinary) Bogomol'nyi point where the equations
to solve reduce, in ordinary space, from second
to first order. One can see that also in the noncommutative
case, this choice allows to write the energy as a sum of squares plus  surface terms,
\be
E =
2\pi \theta {\rm Tr} \left(
\frac{1}{2}\left( B \mp  e(\phi\bar \phi)^2\right)+
2  D_z\phi D_{\bar z} \bar \phi
+ (e\eta^2 B - T)
\right)
\label{lag2}
\ee
We shall not explicitly write $T$ but note that
one can easily prove that Tr$T=0$. Then, one gets a Bogomol'nyi bound in the form
\be
E \geq 2\pi e\theta \eta^2 Tr B
\ee
and the bound is attained when the Bogomol'nyi equations hold,
\ba
B &=& \  e(\eta^2 -\phi\bar\phi ) \; , \;\;\; D_{\bar  z} \phi = 0 \;\;\;\;
{\rm selfdual} \label{equi1}\\
-B &=& e(\eta^2 -\phi\bar\phi ) \; , \;\;\; D_{ z} \bar \phi= 0 \;\;\;\; {\rm anti-selfdual}
\label{equi2}
\ea

In order to find a solution to these equations
(or to the eqs. of motion
were we aot from the Bogomol'nyi point) we can follow the
 three steps in Nielsen-Olesen demarche, which become now
 (we take $N=1$ for simplicity)
\begin{eqnarray}
{\rm  ordinary~space} \hspace{0.2cm}  &\Rightarrow&  \hspace{0.2cm} {\rm
noncommutative~space}
\nonumber\\
 \underbrace{|\phi| = {\eta}}_
{trivial} \hspace{0.2cm}  &\Rightarrow&  \hspace{0.2cm} \phi =
\eta \sum
\underbrace{f^F_n}_{0, \pm 1}|n\rangle \langle n |
\nonumber\\
 \underbrace{|\phi| = \eta \exp(i\varphi)}_
 {singular}= \eta \frac{z}{|z|}
  \hspace{0.2cm} & \Rightarrow &  \hspace{0.2cm}
\phi = \eta \sum
\underbrace{f^F_n}_{0, \pm 1} |n\rangle \langle n | \hat a
\nonumber
\\
\underbrace{|\phi|\! = \!f(r) \exp(i\varphi)}_
{regular}= \!f(|z|)\frac{z}{|z|}
 \hspace{0.2cm} &\Rightarrow&  \hspace{0.2cm}
\phi = \sum
{f_n}|n\rangle \langle n | \hat a
%\nonumber\\
\label{numb}
\end{eqnarray}
Note that in the last two lines we have identified
$z$ with $(1/\sqrt \theta) \hat a$. The difference between the two
expansions is that
in the second  one coefficients $f^F_n$ are are fixed to be $0,\pm 1$
while in the third
one the $f_{n}$  coefficients should be
adjusted using the equations of motion (or Bogomol'nyi equations)
 and boundary conditions.

Of course (\ref{numb}) should be
accompanied by
 a consistent   ansatz for the gauge field.
 One easily finds that the appropriate choice is

\begin{eqnarray}
 \phi&=& \eta \sum_n f_n |n\rangle \langle n+1| \label{an}\\
 A_z&=&\frac{i}{\sqrt \theta} \sum_n
(\sqrt{n+1}-\sqrt{n+2}+e_n)|n+1\rangle \langle n| \label{an0}
\end{eqnarray}
The first two terms in $A_z$ are those arisi9ng for a pure gauge. The coefficients
$e_m$'s should  then be adjusted, together with $f_n$'s, so
that they lead to a non-trivial vortex gauge field
at finite distances and go to 0 at infinity, where $F_{z\bar z} = 0$.
To determine the coefficients $f_n$  and $e_n$, one has
  to   solve the algebraic
equations that arise when plugging expansions (\ref{an})-(\ref{an0}) into equations
(\ref{equi1}) or (\ref{equi2}). To understand how one proceeds, let us
write the Higgs field in (\ref{an}) in the form
\begin{equation}
\hat \phi= \eta \frac{f(\hat N)}{\sqrt{\hat N+1}}
\frac{z}{\sqrt{2\theta}}
\end{equation}
with $\langle n |f(\hat N)|n\rangle=f_n$. In this form,  one
can compare the noncommutative ansatz
with the original Nielsen-Olesen one in
ordinary space,
\begin{equation}
\phi= \eta \,g(|z|) \,z
\end{equation}
In ordinary space the value $g(0)$ at the origin was determined by solving
the Bogomol'nyi
equations and requiring that it adjusts so that at infinity $g(|z|) \sim 1/|z|$.
This implies a numerically subtle process of analytic continuations.
The result is\cite{deVega:1976mi}
\be
g(0)^2=0.72791
\label{nume}
\ee

One can proceed in the same way in the noncommutative case but no
analytic continuations are needed. Remember that
space at infinity corresponds, in the Fock space approach, to large values of $n$.
Then the requirement at infinity $g(|z|) \sim 1/|z|\,, \;\; |z|\gg 1 $ translates into
$ f_n \sim 1  \,, \;\;  n \gg 1$.

As an example, consider the selfdual case. Using  ansatz{\ae} (\ref{an0})-(\ref{an}),
eqs.(\ref{equi1}) become the recurrence equations
\begin{eqnarray}
\sqrt{(n+2)}(f_{n+1}-f_{n})-e_n f_{n+1}&=&0  \nonumber\\
2\sqrt{(n+1)} e_{n-1}-e^2_{n-1}- 2\sqrt{(n+2)} e_{n}-e^2_{n}&=&-
\theta \eta^2 (f_n^2-1) \label{ran}
\end{eqnarray}
  This coupled system
 can be combined to give for $f_n$
\begin{eqnarray}
f^2_{1} &=& \frac{2 f^2_0}{1+\theta \eta^2 - \theta \eta^2
(f_0^2)} \nonumber\\ f^2_{n+1} &=& \frac{(n+2) f^4_n}{f_n^2
-\theta \eta^2 f_n^2(f_n^2-1)+(n+1)f^2_{n-1}} \, \;\;\;\; n >0
\label{rec}
\end{eqnarray}
Following a ``shooting'' procedure, one starts with a given a value for $f_0$
and then determine all $f_n$'s from
(\ref{rec}). The correct value  for $f_0$  should make $f_n^2 \to
1$ asymptotically, so that boundary conditions are satisfied. Of course, the
final correct values for all coefficients will depend on the choice of the
dimensionless parameter $\theta \eta^2$.
For small $\theta$ (commutative limit) one can very easily check
that one re-obtains the
ordinary space values
for the solution. Indeed,
\begin{equation}
\frac{f_0^2}{2\eta^2\theta} =0.72792  \,\,\,\,\,\,\,\,\, \theta<<1
\end{equation}
(compare with eq.(\ref{nume})),
 while for large $\theta$ one reproduces the result of ref.[27].
\begin{equation}
f_0^2=1-\frac{1}{\eta^2\theta} \,\,\,\,\,\,\,\,\, \theta>>1
\end{equation}
Exploring the whole range of $\theta \eta^2$, one finds that
the vortex solution with $+1$ units of magnetic flux  exists  in
all the intermediate range.
 As an example,
we list three representative values,
\begin{eqnarray}
\theta \eta^2 = 0.5, & & f_0^2 = 0.40069\ldots \nonumber\\
\theta \eta^2 = 1.0, & & f_0^2 = 0.56029\ldots \nonumber\\
\theta \eta^2 = 2.0, & & f_0^2 = 0.70670\ldots
\end{eqnarray}
Once all $f_n's$ and $e_n's$ are calculated, one can compute the
magnetic field,  for example from the formula
\be
\hat B = \eta^2 \sum_{n=0}^{\infty} \left(1 - f_n^2\right) \vert n
\rangle \langle n \vert \label{ayu} \ee
Now,  the explicit formula for $\vert n \rangle \langle  n
\vert $ in configuration space\cite{GMS} allows to write
\be
B(r) = 2 \eta^2 \sum_{n=0}^{\infty}\!(-1)^n \left(1 -
f_n^2\right) \exp\left({-\frac{r^2}{\theta}}\right)
 L_n(\frac{2r^2}{\theta}) \; , \;\;\;
 \Phi = 2 \pi \theta {\rm Tr} \hat B = 2\pi
 \ee
\begin{figure}
\centerline{\psfig{file=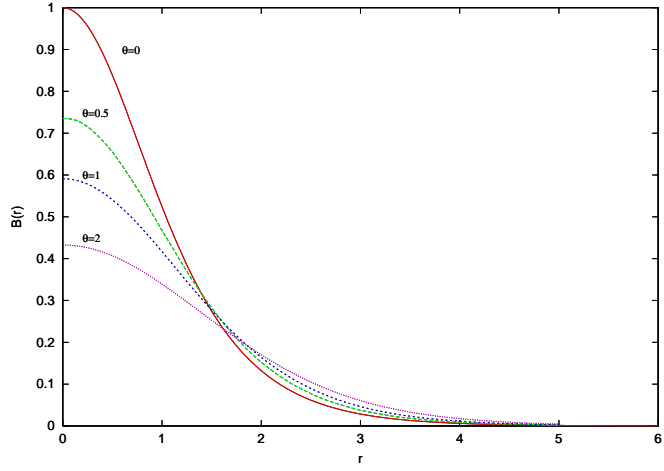,width=9cm,angle=-90}}
\vspace*{8pt} \caption{Magnetic field of a self-dual vortex as a
function of the radial coordinate (in units of $\eta$) for
different values of the anticommuting parameter $\theta$ (in units
of $\eta^2$). The curve for $\theta = 0$ coincides with that of
the Nielsen-Olesen vortex in ordinary space. \label{fig1}}
\end{figure}

We show in Figure \ref{fig1} the resulting magnetic field
$B$ as a function
of $r\eta$. As explained above,  one recovers for $\theta = 0$ the result
for self dual
Nielsen-Olesen vortices in ordinary space\cite{deVega:1976mi}. As $\theta$
grows, the maximum for B decreases and the vortex is less
localized with total area such that the magnetic flux remains
equal to 1. It is important to stress that one finds
noncommutative \underline{self-dual vortex} solutions in the whole range of
$\theta$, in agreement with the analysis for large and small
$\theta$ presented in [27].

In ordinary space, anti-selfdual (negative flux) solutions   can be trivially
obtained  from selfdual ones, just by making $B \to -B$,
 $\phi \to \bar\phi$.
Now, the presence of  the noncommutative parameter $\theta$,
breaks parity and the moduli space for positive
and negative magnetic flux vortices differ drastically.
One has then to carefully study this issue in all regimes, not only
for $\lambda = \lambda_{BPS}$ but also for
$\lambda
\ne \lambda_{BPS}$, when Bogomol'nyi equations do not hold
and  the second order equations
of motion should be analyzed. Of course the recurrence
relations associated to the second-order equations of motion become more
complicated than (\ref{equi1})-(\ref{equi2}); for example, for positive flux,
they read
\begin{eqnarray}
2( t_{n} f_{n+1}\; \sqrt{n+1+M} + t_{n-1}f_{n-1}\; \sqrt{n+M}
)+&&\nonumber\\
&&\hspace{-6cm}(t^2_{n}+t^2_{n-1}+2n+2M+1)f_n =-
\frac{{\theta{\eta}^{2}}{\lambda}}{2} f_n({f_{n}}^{2}-1) \; , \;\; n>1
\nonumber
\end{eqnarray}
\[
(t_{n+1}^2-2 t^2_{n}+t_{n-1}^2)t_n
={\theta}{\eta}^{2}\left(2f_{n}f_{n+1}\;{\sqrt{n+1+M}}
 +({f_{n}}^{2}
+{f_{n+1}}^{2})t_n \right)  \; , \;\; n>1
\]
\begin{eqnarray}
&&f_{1}= -\frac{f_0}{2t_0 \sqrt{1+M}} \left(
 (1+2M) + t_{0}^2 +
\frac{\theta\eta^{2}\lambda}{2}({f_{0}}^{2}-1)\right)\nonumber\\
&& t_1^2=2t_0^2 + \theta \eta^2 \left( (f_1^2+f_0^2) t_0 + 2
\sqrt{1+M} f_0 f_1/t_0  \right) \label{tres}
\end{eqnarray}
where we have slightly changed  the definition of the gauge field
coefficients just for notational
convenience,
\begin{eqnarray}
\hat{\phi} &=&\eta\sum_n f_{n}|n\rangle \langle n+M|\nonumber\\
{\hat A}_z &=&\frac{i}{\sqrt\theta} \sum_n (t_{n}
+\sqrt{n+1})|n+1\rangle \langle n| \label{ansatz2}
\end{eqnarray}
One can still solve this equations as for the simpler BPS case.
We give below a summary of
the main results\cite{LMS1,LMS2,LMRS}. As before, we take $\theta >0$ and
distinguish positive from negative magnetic flux (this being
equivalent to a classification of selfdual and anti-selfdual solutions).

~

\begin{itemize}
\item {\bf Positive flux}
\begin{enumerate}
\item  {There are BPS and non-BPS solutions in
the whole range of
$\eta^2\theta$. Their energy and magnetic flux are:

\noindent For BPS solutions
\begin{eqnarray}
E_{BPS} &=& 2\pi \eta^2 N \; , \;\;\;    \Phi = 2 \pi N
\nonumber\\
N &=& 1,2,\ldots
\end{eqnarray}

 For non-BPS solutions,
\begin{eqnarray}
E_{non-BPS} &>& 2\pi \eta^2 N \; , \;\;\;    \Phi = 2 \pi N
\nonumber\\
N &=& 1,2,\ldots
\end{eqnarray}
\item For $\eta^2\theta \to 0 $   solutions
 become,
smoothly, the known regular solutions of the commutative case.
\item
In the non-BPS case, the energy of an $N=2$
vortex compared to that of two $N=1$ vortices
is a function of $\theta$.}
\end{enumerate}
As in the commutative case, if one compares the energy of an $N=2$
vortex  to that of two $N=1$ vortices  as a function of
$\lambda$ one finds that for $\lambda> \lambda_{BPS}$ $N>1$ vortices are
unstable -they repel- while for  $\lambda < \lambda_{BPS}$ they
attract.
%\begin{figure}%[ht]
%\centerline{
%\psfig{figure=energyp2.ps,height=5.5cm,width=7.5cm,angle=0}}
%\smallskip
%\caption{ The energy for an ${N=2}$ vortex (solid line)
%compared to that of two $N=1$ vortices (dashed line)
%as a function of $\lambda$, for $\theta\eta^2 = 2$.}
 %\label{fig2}
%\end{figure}

~

\item {\bf Negative flux}
\begin{enumerate}
\item
 BPS  solutions only exist
in a finite range:
\[0 \leq \eta^2\theta \leq 1\]
Their energy and magnetic flux are:
\begin{eqnarray}
E_{BPS} &=& 2\pi \eta^2 N \; , \;\;\;    \Phi = 2 \pi N  \nonumber\\
N &=& 1,2,\ldots
\end{eqnarray}
\item
When $\eta^2\theta = 1$ the BPS solution becomes a {fluxon},
a configuration which is regular only in the noncommutative
case. The magnetic field of a typical fluxon solution takes the form
\be
B
\sim \frac{1}{\sqrt \theta}{\exp (r^2/\theta)}
\ee
We stressed above that for  $\theta \to 0$ such a  configuration becomes
singular but note  that in the present case the noncommutative
parameter is fixed, $\theta = 1/\eta^2 \ne 0$.
\item
There exist non-BPS solutions in the
whole range of $\theta$ but
\begin{enumerate}
\item Only for $\theta <1$
they are smooth deformations of the
commutative ones.
\item For   $\theta \to 1$ they
tend to the fluxon BPS solution.
\item For  $\theta >1$ they coincide
with the non-BPS fluxon solution.
\end{enumerate}
\end{enumerate}
\end{itemize}

\section*{Noncommutative instantons}
The well-honored  instanton equation
\be
F_{\mu \nu} = \pm \tilde F_{\mu\nu}
\label{instanton}
\ee
was studied in the noncommutative case by
Nekrasov and Schwarz\cite{NS} who showed that
even in the $U_*(1)$ case one can find nontrivial instantons. The
approach followed in that work was the extension of the
ADHM construction, successfully applied to the systematic construction
of instantons in ordinary space, to the noncommutative case. This
and other
approaches were discussed
in [39]-[46]. Here we shall describe the methods developed in
[15]-[16].

We work in  four dimensional space where one can always choose
\begin{eqnarray}
\theta_{12} = - \theta_{21} = \theta_1 \nonumber\\
\theta_{34} = - \theta_{43} = \theta_2 \nonumber\\
{\rm all ~other~} \theta's = 0 \nonumber
\end{eqnarray}
We define dual tensors as
\be
\tilde{F}_{\mu\nu} = \frac{1}{2} \sqrt g \, \varepsilon_{\mu\nu\alpha\beta}
F^{\alpha\beta}
\ee
with $g$ the determinant of the metric.

In order to work in Fock space as we did in the case of
noncommutative vortices, we now need
two pairs of creation
annihilation operators,
\begin{eqnarray}
x^1 \pm i x^2 ~ ~ ~ &{\Rightarrow}& ~ ~ ~ {\hat a}_1, \;\;
{\hat a}^{\;\dagger}_1 \nonumber\\
x^3 \pm i x^4 ~ ~ ~ &{\Rightarrow}& ~ ~ ~ {\hat a}_2, \;\;
{\hat a}^{\;\dagger}_2 \nonumber
\end{eqnarray}
and  the
Fock vacuum will be denoted as $| 0 0 \rangle$. Concerning
projectors  the connection with configuration space takes the form
\begin{eqnarray}
|n_1 n_2\rangle \langle n_1 n_2| &{\Rightarrow}&
\exp\left( -r_1^2/\theta_1 -  r_2^2/\theta_2\right)
L_{n_1}\left(2r_1^2/\theta_1\right)\times \nonumber\\
&& L_{n_2}\left(2r_2^2/\theta_2\right)
\label{suis}
\end{eqnarray}
Finally, note that  the gauge group $SU(2)$ (for which ordinary
instantons were originally constructed) should be replaced
 by $U(2)$ so that
\be
A_\mu = A_\mu^a \frac{\sigma^a}{2}
 + A_\mu^4 \frac{I}{2}
\ee

Let us now analyze how the different ansatz leading
to ordinary instantons can be adapted to the noncommutative case.

~

\noindent {\bf i- (Commutative)'t Hooft multi-instanton ansatz}
(1976)

\vspace{0.2 cm}

Gauge fields take value in the Lie algebra of $SU(2)$. They are written as
\begin{eqnarray}
&& A_\mu(x) = \tilde{\Sigma}_{\mu\nu} j_\nu
\nonumber\\
&&
\tilde{\Sigma}_{\mu\nu} = \frac{1}{2}\bar\eta_{a\mu\nu}
 \sigma^a \; , \;\;\;
\bar \eta_{a\mu\nu} =
\left\{
\begin{array}{l}
 ~\varepsilon_{a\mu\nu} \; , \;\;\; \mu,\nu \ne 4\\
  ~~\delta_{a\mu} \; , \;\;\; \nu=4\\
 -\delta_{a\nu}\; , \;\;\;\mu=4
\end{array}
\right.
\nonumber
\\
&&
j_\nu = \phi^{-1} \partial_\nu \phi \nonumber
\end{eqnarray}
Here $\sigma^a$ are the Pauli matrices. With this ansatz, one
can prove that
\be
F_{\mu \nu} = \tilde F_{\mu\nu}
+
\frac{1}{\phi} \nabla \phi
\label{lapl0}
\ee
so that if one can find a solution of the equation
\be
\frac{1}{\phi} \nabla \phi = 0
\label{lapl}
\ee
it will lead to a selfdual instanton configuration. The proposal
of 't Hooft was to take $\phi$ as
\begin{eqnarray}
\phi = 1 + \sum_{i=1}^N\frac{\lambda_i^2}{(x - c_i)^2}
\label{sesenta0}
\end{eqnarray}
with $c_i = (c_i^\mu)$ and $\lambda_i$ real constants.
Such a $\phi$ satisfies
\be
\nabla \phi = \sum_{i=1}^N\delta^{(4)}(x - c_i)
\ee
Now,  each  $\delta$-function singularity in $\nabla \phi$ is cancelled
in  (\ref{lapl}) by
the corresponding one in $\phi$ so that the equation is satisfied everywhere
and selfduality in (\ref{lapl0}) is achieved.
The solution   corresponds to a regular
 instanton of topological charge $Q=N$. Points $c_i$ can be interpreted
 as the centers of spheres of radius $\lambda_i$ where
 one unit of topological charge is concentrated. One should note
 that the gauge field has   singularities at points $x^\mu = c_i^\mu$
 but these singularities are gauge
 removable and so the field strength
is everywhere regular. For example, in the $Q=1$ case,
the gauge field associated to 't
 Hooft ansatz reads
 \be
 A_\mu = -2 \lambda^2 \bar \Sigma_{\alpha\beta}
 \frac{x_\beta}{x^2(x^2 + \lambda^2)}
 \label{ssin}
\ee
Since we have chosen $c_1 =0$, the singularity is located at the origin.
A (singular) gauge transformation leads to  the regular
version of (\ref{ssin}),
\be
 A_\mu = -2  \Sigma_{\alpha\beta}
 \frac{x_\beta}{(x^2 + \lambda^2)}
 \label{rrin}
\ee

\noindent {\bf ii- Noncommutative version of 't Hooft  ansatz}

\vspace{0.2 cm}

To find a 't Hooft-like ansatz for  noncommutative instantons, one not only has
to  extend the proposal in the $SU(2)$
sector to the noncommutative case  but one has also to include an
ansatz for the $A_\mu^4$ component.  A judicious choice is
\begin{eqnarray}
&& {A_\mu^a \frac{\sigma^a}{2} =
 \bar \Sigma_{\mu\nu} j_\nu}  ~ ~ ~ \Rightarrow  ~ ~ ~
 A_\mu^a \frac{\sigma^a}{2} =
 \bar \Sigma_{\mu\nu} j_\nu \, , \;\;\; a=1,2,3
\label{juta}\\
&& j_\nu = \phi^{-1} \partial_\nu \phi  ~ ~ ~  \Rightarrow ~ ~ ~
j_\nu = \Phi^{-1} * \partial_\nu \Phi +
\partial_\nu \Phi * \Phi^{-1}
\label{ahora}\\
~ \nonumber
\\
&& A_\mu^4 =
-i \left(\Phi^{-1} * \partial_\nu \Phi -
\partial_\nu \Phi * \Phi^{-1}\right)
\label{jita}
\end{eqnarray}
The extension in (\ref{juta})-(\ref{ahora}) is the most natural one: ansatz (\ref{juta}) is
just the one proposed by 't Hooft and (\ref{ahora}) is the symmetrized
version, with respect to the $*$-product, of 't Hooft current for ordinary
products (Note that in the commutative limit $\Phi^2 = \phi$).
Concerning the completely new expression
(\ref{jita}), it is the simplest ansatz, compatible with (\ref{juta}).
In the commutative limit it corresponds to $A_\mu^4 = 0$ and, more important,
it leads to a selfdual  $F_{\mu\nu}^4$. This is not a trivial
fact since   $F_{\mu\nu}^4$ picks
contributions not only from $A_\mu^4$ but also from  $A_\mu^a\,, \; a=1,2,3$)
but, strikingly,  (\ref{jita}) ensures selfduality of $F_{\mu\nu}^4$ for
\underline{any} $\Phi$.

Now, one also needs selfduality of
$F_{\mu\nu}^a$ with $a=1,2,3$. As in the case of $F_{\mu\nu}^4$, also the
$SU(2)$ field strength picks contribution from $A_\mu^4$. After some work one can
prove that
\be
\tilde F_{\mu\nu}^a = F_{\mu\nu}^a  + \eta_{\mu\nu a}
 \Phi^{-1} * \nabla \Phi^2 * \Phi^{-1}
\ee
In the ordinary case, finding a solution of $\nabla \phi$
(which translates into $\nabla \Phi^2$ in the present case) was
enough to ensure selfduality everywhere. Inspired in this
fact, and noting that, as  we learnt when studying fluxons
one has, on the one hand
\be
\lim_{\theta \to 0} \frac {1}{\theta} \exp\left(-\frac{r^2}{\theta}\right)
\sim \delta(\vec r)
\ee
and, on the other,
\be
|00><00| \sim  \frac {1}{\theta_1\theta_2}
\exp\left(-\frac{r_1^2}{\theta_1} -
\frac{r_2^2}{\theta_2}\right)
\ee
we can try to pose the problem in the form
\begin{eqnarray}
&& \frac{1}{\phi} \nabla \phi = 0 ~ ~ ~  \Rightarrow  ~ ~ ~
\Phi^{-1} *\nabla \Phi^2 *\Phi^{-1} = 0 \nonumber\\
&&  \nabla \phi = \delta^{(4)}(x)
 \Rightarrow  ~ ~ ~
 \nabla \Phi^2 =  \frac{\lambda^2}{\theta_1\theta_2}|00\rangle\langle00|
 \label{sssdd}
\end{eqnarray}
One easily finds the solution to the equation
in the r.h.s. of (\ref{sssdd}). It reads   (for simplicity we consider
here the $N=1$ case and put  $\theta_1=\theta_2 = \theta$)
\be
\Phi^2 = 1 + \frac{\lambda^2}{r_1^2 + r_2^2}\left(
1 - \exp\left(-\frac{r_1^2 + r_2^2}{\theta}\right)
\right)
\label{version}
\ee
where, say, $\vec r^1 = (x^0,x^1)$ and $\vec r^2 = (x^2,x^3)$.
The  Fock space version of (\ref{version}) is
\be
\Phi^2  = 1 + \frac{\lambda^2}{2\theta}\sum \frac{1}{n_1 + n_2 + 1}
|n_1n_2\rangle \langle n_1 n_2|
\ee
Now, in contrast with what happened
in ordinary space, where $\phi^{-1} \nabla \phi = 0$  one    gets now
\be
\Phi^{-1} \nabla \Phi^2 \Phi^{-1} = -\frac{2\lambda^2}{\theta(2\theta + \lambda^2)}
|00\rangle\langle 00|
\ee
so that the field strength and its dual satisfies  a relation of the form
\begin{eqnarray}
 && F_{\mu \nu} = \tilde F_{\mu \nu}  +
 \frac{2\lambda^2}{\theta(2\theta + \lambda^2)}\tilde \Sigma
|00\rangle\langle 00|
\label{last}
\end{eqnarray}
We see   that the
self-dual equation is not exactly satisfied: the $|00\rangle\langle00|$
term, the analogous to the delta function in the
ordinary case, is not cancelled as it happened
 with the delta function
source for the Poisson equation (\ref{lapl}) in the commutative case.

~

\noindent{\bf iii- Noncommutative BPST ($Q=1$) ansatz} (1975)

~

The pioneering Belavin, Polyakov, Schwarz and Tyupkin
ansatz\cite{BPST} leading
to the first $Q=1$ instanton solution was similar to the
't Hooft ansatz except that $ \Sigma_{\mu\nu}$ was used instead
of its dual $\tilde{ \Sigma}_{\mu\nu}$. Its noncommutative extension
can be envisaged by writing, in the $SU(2)$ sector,
\begin{eqnarray}
& & {A_\mu^a \frac{\sigma^a}{2} =
 \Sigma_{\mu\nu} j_\nu} ~ ~ ~   \Rightarrow  ~ ~ ~
A_\mu^a \frac{\sigma^a}{2} =
 \Sigma_{\mu\nu} j_\nu
\label{aotro}
\end{eqnarray}
where $j_\nu$ is defined as in the previous ansatz. Concerning $A_\mu^4$, the consistent
ansatz changes due to the use of $\Sigma_{\mu\nu}$ instead of its dual
as done in  the 't Hooft ansatz.
One needs now,
instead of (\ref{jita}),
\begin{eqnarray}
& & A_\mu^4 =
i \left(\Phi^{-1} * \partial_\nu \Phi + 3
\partial_\nu \Phi * \Phi^{-1}\right)
\label{a4}
\end{eqnarray}
With this, one finally has
\begin{eqnarray}
 & & F_{\mu\nu} = \tilde F_{\mu \nu}\; , \;\;\;    Q = S = 1
\end{eqnarray}
but, because of the necessity of a consistent ansatz for the $A_\mu^4$
component forces a factor ``3'' in the second term in the
r.h.s. of eq.(\ref{a4}), one can see that
\be
F_{\mu\nu} \ne  F_{\mu \nu}^\dagger
\ee
and hence the price one is  paying in order to have a selfdual field
strength is its  non-hermiticity.
Note however that the action and the topological charge are real.

~

\noindent{\bf iv- (Commutative) Witten ansatz}\cite{Witi} (1977)

~

The clue in this ansatz
is to reduce the four dimensional problem to
a two dimensional one through an
 axially symmetric N-instanton ansatz. That is, one passes from
 $d=4$   Euclidan space to $d=2$ space with non trivial metric,
$(x^1,x^2,x^3,x^4) \to (r,t) $.

The axially symmetric ansatz for the gauge field components
is
\begin{eqnarray}
\vec{A}_r  & = &  A_r(r,t) \vec \Omega(\vartheta,\varphi)  \nonumber\\
\vec{A}_t  & = &  A_t(r,t) \vec \Omega(\vartheta,\varphi) \nonumber\\
\vec A_\vartheta &=& \phi_1(r,t) \partial_\vartheta \vec
\Omega(\vartheta,\varphi)
+ \left(1 + \phi_2(r,t)\right) \vec \Omega (\vartheta,\varphi)
 \wedge \partial_\vartheta \vec \Omega (\vartheta,\varphi)
\nonumber\\
\vec A_\varphi &=& \phi_1(r,t) \partial_\varphi
 \vec \Omega(\vartheta,\varphi)
 +
\left(1 + \phi_2(r,t)\right) \vec
\Omega (\vartheta,\varphi) \wedge \partial_\varphi \vec \Omega
(\vartheta,\varphi)
\label{ansatz}
\end{eqnarray}
with
\begin{equation}
\vec \Omega(\vartheta, \varphi)  = \left(
\begin{array}{c}
\sin \vartheta \cos \varphi \\
\sin \vartheta \sin \varphi \\ \cos \vartheta
\end{array}
\right)
\ee
With this ansatz the Yang-Mills action reduces
to an Abelian-Higgs action in a curved space with metric
\be
g^{ij} = r^2 \delta^{ij}
\ee
\begin{eqnarray}
\frac{1}{4}{\rm Tr} \int d^4 x F_{\mu\nu}F_{\mu\nu} &=& \int_{-\infty}^\infty dt
\int_0^\infty dr \sqrt g \left(\frac{1}{2} g^{ij}D_i\phi^a D_j\phi^a +
\frac{1}{8} g^{ip}g^{jq}F_{ij}F_{pq} + \right.\nonumber\\
&& \left. \frac{1}{4}(1 - \left(\phi^1)^2 + (\phi^2)^2\right) \right)
\end{eqnarray}
The
 selfduality instanton equations (\ref{instanton}) associated to the Yang-Mills action
become a pair of BPS equations  for the gauge field -Higgs action
in curved space,
\begin{eqnarray}
F_{\mu\nu} = \tilde F_{\mu\nu} &\to &
\left\{
\begin{array}{l}
\frac{1}{\sqrt g}F_{z \bar z} = |\phi|^2 - 1\\
D_z \phi = 0
\end{array}
\right.
\end{eqnarray}
where $\phi = \phi_1 + i \phi_2$ and $z = t + i r$.
Moreover, solving these
  BPS vortex equations can be seen to reduce to finding the
  solution of a Liouville equation. In this way an
exact axially symmetry N-instanton solution was
constructed\cite{Witi} for the (commutative) $SU(2)$ theory.

~

\noindent {\bf v- Noncommutative version of Witten ansatz}

As we learnt in lecture 1, given a commutation relation of the form
\be
[x^i,x^j] =  \theta^{ij}(x)
\ee
not all functions $\theta_{ij}(x)$ will guarantee a
noncommutative but associative product.
In the present 2 dimensional case we showed that a sufficient condiction
for this was given by eq.(\ref{clase1}),
\be
\theta_{ij} = \theta_0 \frac{\varepsilon_{ij}}{\sqrt g}
\ee
with $\theta_0$ a constant. Then, given the metric in
which the instanton problem with axial symmetry reduces to
a vortex problem we see that an associative  noncommutative
product should take the form
\be
[r,t] = i r^2 \theta_0
\label{rara}
\ee
with now $r$ and $t$ defining the two dimensional variables
in curved space. We then take this as the non-trivial commutation relation
in the four-dimentional space from which the two-dimensional problem is
inherited and put all the other commutation relations to zero,
\be
[r,\vartheta] = [r, \varphi] = [\vartheta,\varphi] = [t,\vartheta]=[t, \varphi]=0
\ee

A further simplification occurs after the observation that
\be
 r*t - t* r = r^2 \theta_0  ~ ~ ~
\Rightarrow ~ ~ ~ t   * \frac{1}{r} - \frac{1}{r} * t = \theta_0
\ee
Then, calling $y^1 = t$ and $y^2 = 1/r$ we have
instead of (\ref{rara})
the usual flat space Moyal
product and the  Bogomol'nyi equations take the form
\begin{eqnarray}
 \left( 1 - \frac{1}{2} (z + \bar z )^2\right)D_z \phi &=&
 \left( 1 + \frac{1}{2} (z + \bar z )^2\right)D_{\bar z} \phi
\label{cuss} \\
iF_{z \bar z} & = & 1 - \frac{1}{2}[\phi,\bar \phi]_+
\label{cus}\\
iF_{  z \bar z} & = & - \frac{1}{2}[\phi,\bar \phi] \label{cu}
\end{eqnarray}
with $z = y^1 + i y^2$. We can at this point apply the
Fock space method detailed above for constructing vortex solutions.
In the present case, consistency of eqs.(\ref{cuss})-(\ref{cu}) imply
\be
\bar \phi \phi =1
\ee
and hence the only kind of nontrivial ansatz
should lead, in Fock space, to a scalar field of the form
\be
\phi = \sum_{n=0} |n + q\rangle \langle   n|
\ee
where $q$ is some fixed positive integer.
With this, it is easy now to construct a class of solutions analogous
to those discussed in the context of fluxons
  in
flat space. Indeed, a configuration of the form
\begin{eqnarray}
\phi &=& \sum_{n=0}|n+q\rangle\langle n|  \nonumber\\
A_z &=& -\frac{i}{\sqrt{\theta_0}} \sum_{n=0}^{q-1} \left( \sqrt{n+1}
\right) |n+1\rangle\langle n|\label{solucion2}
+\frac{i}{\sqrt{\theta_0}} \sum_{n=q} \left( \sqrt{n+1-q} - \sqrt{n+1}
\right) |n+1\rangle\langle n|\nonumber\\
\label{solucion}
\end{eqnarray}
 satisfies eqs.(\ref{cuss})-(\ref{cu})
 provided
 $\theta_0 = 2$. In
 particular, both the l.h.s. and r.h.s of eq.(\ref{cuss})
 vanish separately.
The field strength associated to this solution reads,
 in Fock space,
\be
i F_{z \bar z} = -\frac{1}{2} \left(
|0\rangle\langle 0| +   \ldots +
|q-1\rangle\langle q-1|
\right) \equiv B
\label{defB}
\ee
or, in the original spherical coordinates
\begin{eqnarray}
\vec F_{tu} &=& B(r) \vec \Omega \nonumber\\
\vec F_{\vartheta\varphi}
&=& B(r)   \sin \vartheta \, \vec \Omega \nonumber\\
F_{tu}^4 &=& B(r) \nonumber\\
 F_{\vartheta\varphi}^4 &=& B(r) \sin \vartheta
\label{instan}
\end{eqnarray}
As before, starting from  (\ref{defB}) for $B$ in Fock space, we can
obtain the explicit form of $B(r)$ in configuration space  in terms
of Laguerre polynomials, using eq.(\ref{suis}).
Concerning the topological charge, it
is then given by
\begin{eqnarray}
Q &=& \frac{1}{32 \pi^2} {\rm tr} \int d^4 x {\varepsilon}^{\mu \nu \alpha \beta}
 F_{\mu\nu} F_{\alpha \beta}
= \frac{1}{\pi}
\int_{-\infty}^{0}du \int_{-\infty}^{\infty} dt B^2 = 2 {\rm Tr} B^2
= \frac{q}{2}
\end{eqnarray}
We thus see that $Q$ can be in principle integer or semi-integer,
 and this for an ansatz
which is formally the same as that proposed in [48] for ordinary
space and which yielded in that case to an integer. The origin of this
difference between the commutative and the noncommutative
cases can be traced
back to the
fact that in the former case,  boundary conditions were imposed on the
half-plane and
forced the solution to have an associated integer number.
  In fact, if one plots Witten's
 vortex solution in ordinary space in
the whole $(r,t)$ plane, the magnetic flux has
two peaks and the corresponding
vortex number is even. Then, in order to parallel this treatment in the
noncommutative case one should impose the condition $q = 2 N$.

\subsection*{Monopoles}
Noncommutative monopoles cannot be found as simply as vortices or instantons
 by working in the
Fock space framework. This is because creation-annihilation operator appear by pairs
and monopole configurations should depend on 3 (not on 2 or 4)  variables. However,
it is known since the work of Manton\cite{manton} (see also [51]) that
monopoles can be obtained as a particular limit of instanton solutions. The
basic idea is the following: the instanton equation in 4-dimensional Euclidean space,
\be
 F_{\mu\nu} = \tilde  F_{\mu\nu}
 \label{muno}
 \ee
can be brought into the BPS monopole equation in 3-dimensional space,
\be
F_{ij}^a = \varepsilon_{ijk} (D_k\phi)^a
\label{mdos}
\ee
by identifying
\be
A_0^a = \phi^a
\label{idea}
\ee
and eliminating in some way the dependence on Euclidean time. The connection
between (\ref{muno}) and (\ref{mdos}) is evident. Indeed, for space-space indices,
 (\ref{muno}) reads
\be
F_{ij} = \varepsilon_{ijk} F_{k0}
\ee
Now, if gauge fields were time in dependent and moreover, we identify $A_0$
with $\phi$ according to (\ref{idea}), one gets (\ref{mdos}).

The idea of Manton\cite{manton} was to start from Witten $N$-instanton solution which,
as we have seen, is a sort of superposition of $1$-instantons located in
different points of the time axis. Then, taking the $N \to \infty$ limit wipes out the
dependence on time and hence in this limit the instanton configuration becomes
time independent. One then has the desired connection,
\be
\lim_{N\to \infty} (A_i^{N-inst}(\vec x, t),A_0^{N-inst}(\vec x, t)) \;\;\; \Longrightarrow
\;\;\;
(A_i^{mon}(\vec x), \phi^{mon}(x))
\ee
This $N \to \infty$ has to be taken with care. Remember that the N-instanton was
constructed by solving a vortex (Liouville) equation whose solution has poles
located at certain points that we call $a_i$. In fact, the fundamental function from
which the solution is constructed is
\be
g(z) = \prod_{i=1}^N \frac{a_i - z}{\bar a_i + z}
\ee
The appropriate $N\to\infty$ limit should be taken for all points $a_i$ coinciding
and taking the form
\be
a_i = \frac{2N}{\beta}
\ee
with $\beta$ some scale so that
\be
g(z) = \lim_{N \to \infty} \left(
\frac{1- \beta z/2N}{1+ \beta z/2N}\right) = \exp(-\beta z)
\ee

Let us take the same route but starting from our noncommutative solution. Rememeber that
 the magnetic
field of the $d=2$ vortex in curved space was (\ref{defB})
\be
B^q =  = -\frac{1}{2} \left(
|0\rangle\langle 0| +   \ldots +
|q-1\rangle\langle q-1|
\right)
\label{defBB}
\ee
In the infinite ($q \to \infty$) limit, one clearly has
\be
\lim_{q \to \infty}B^q = \frac{1}{2}
\ee
which is realized by the infinite charge limit of $A_z$ or
\be
\lim_{q \to \infty} A_{1} = -\frac{y_2}{2} \;, \;\;\;
\lim_{q \to \infty} A_{2} = \frac{y_1}{2}
\ee
In order to convert the self-duality instanton into a BPS monopole one has
to identify $A_{2} = \phi$ and $(A_{1},A_\varphi A_\vartheta)$ with the spatial
components of the monopole. But one needs time-independence ($y_2$ independence)
of
the resulting fields and   $A_{1}$ depends on time. In the ordinary
case, one can easily gauge away this time dependence. Here, the affair
becomes more involved:  one has that $\theta =2$  and
then one is in
the peculiar situation discussed at the end of lecture 1 where
gauge orbits could consist in just one point. Indeed, given the gauge field
\be
A_i = \frac{1}{2} \varepsilon_{ij} y_j
\ee
a general gauge transformation takes the form
\be
A_i^g = A_i + \left(1 - \frac{\theta}{2}\right) g^{-1}*\partial_i g = A_i
~~{\rm for}~ \theta = 2
\ee
One can however proceed to make a singular gauge transformation leading to the obvious
time independent field configuration
\be
A_2 = -y_1 \; , \;\;\; A_2 = 0
\ee
The transformation is
\be
g = \exp_x(ic\Lambda)
\ee
with
\be
\Lambda = \frac{1}{2} ( \vec \sigma \cdot \vec \Omega + I) \; , \;\;\; c =
\frac{1}{4}\log\left(1 + \frac{\theta}{2}\right)
\ee
Calling $B_i$ the gauge transformed fields, we have that in the new gauge the only
non-trivial fields are
\begin{eqnarray}
{\vec B}_\varphi &=& \frac{1}{2}(1 + \cos \vartheta)\vec \Omega - \vec \Omega \wedge
\partial_\varphi \vec \Omega\nonumber\\
B_\varphi^4 &=& -  \frac{1}{2}(1 + \cos \vartheta) \nonumber\\
\vec \phi &=& \frac{B}{r} \Omega\nonumber\\
\phi^4 &=& \frac{B}{r}
\end{eqnarray}
Such a gauge field corresponds to a Wu-Yang  monopole\cite{wu-yang} (the non-Abelian
version of the Dirac monopole) coupled to a Higgs scalar. It satisfies
the Bogomol'nyi equation,
\be
\frac{1}{2} \varepsilon^{ijk}
\left(\partial_j A_k - \partial_k A_j +[A_j,A_j]\right) = D^i\phi
\ee
In particular, if we define the ``electromagnetic'' field strength as usual,
\be
{\cal F}_{jk} = {\rm Tr} \frac{\phi}{|\phi|}\left(\partial_j A_k - \partial_k A_j +[A_j,A_j]
\right)
\ee
we get the magnetic field off a Dirac monopole with unit charge,
\be
{\cal B}^r = \frac{\varepsilon^{r \vartheta \varphi}}{\sqrt g}
{\cal F}_{\vartheta\varphi} = -\frac{1}{r^2}
\ee
Of course,  the energy associated to this solution,
 \be
 E =  {\rm Tr} \int d^3x \left( D_i\Phi
D_i\Phi + \frac{1}{2}F_{ij}
 F_{ij} \right) \label{sees} \ee
is strictly infinite (as it coincides with the selfenergy of a
Dirac monopole)
\be
E = \pi  \int dr \frac{1}{r^2} =  \int d^3 x B_{mon}^2
\ee
Now, if
we introduce a regulator $\epsilon$\footnote{Regulator $\epsilon$
 is dimensionless
since $r$ is a dimensionless variable.} to cut off the short-distances
divergence and
recover the dimensional scale $R$ ($\theta =\theta_0 R^2 = 2 R^2$) we
can write $E$ in the form
\be
E =  \frac{\pi}{g_{YM}^2 R\epsilon} =  \frac{\pi R }{g_{YM}^2 R^2 \epsilon} =
\frac{2\pi}{g_{YM}^2\theta} \frac{R}{\epsilon}
\label{E}
\ee
(We have reintroduced the gauge coupling constant $g_{YM}$ which was taken
 equal to 1 along the paper). Defining a length $L = R/\epsilon$
we see that $E$ can be identified with the mass of a string of
length $L$  whose tension is
\be
T = \frac{2\pi}{g_{YM}^2\theta}
\label{T} \ee
One can see (\ref{sees}) as emerging in the decoupling linearized
limit of a $D3$-brane in the Type IIB string theory with the Higgs
field describing its fluctuations in a transverse
direction. Since the
$B$-field leading to our noncommutative setting is transverse to
the $D3$-brane surface, one can make an analysis similar to that
presented by Callan-Maldacena\cite{CM} with the scalar field
describing a perpendicular spike. In this last investigation,
where the electric case is discussed, the string interpretation
corresponds to an $F$-string attached to a $D3$-brane. Our
magnetic case can be related to this by  an $S$-duality
transformation changing the $F1$ into a $D1$ string. Comparing the
tension of such a $D1$-string with the one resulting from our
solution (eq.(\ref{T})),
\be
T_{D1} = \frac{1}{2\pi \alpha'g_s} = \frac{2\pi}{g_{YM}^2\theta}
\ee
and  using $2\pi g_s = g^2_{YM}$ we see that quantization of the
magnetic monopole charge leads to a quantized value  for $\theta$
in string length units equal to $1$ for our charge-1 monopole,
$\theta/2\pi \alpha' = 1$.

 \section{Noncommutative Theories in d=2,3 dimensions}

\subsection*{The Seiberg-Witten map}
This is an explicit map connecting a given noncommutative gauge
theory with a conventional gauge theory.
 Consider the case in which the noncommutative gauge
theory is governed by a Yang-Mills (YM) Lagrangian for the gauge potential
$\hat A_\mu$, transforming under gauge rotations according to
\be
\hat \delta_{\hat \varepsilon} A_\mu(x) =
\hat A'_\mu(x) - \hat A_\mu(x) =   D_\mu[\hat A]\hat \varepsilon(x)
\ee
The Seiberg-Witten map
connects the noncommutative YM Lagrangian
 to some unconventional Lagrangian on the commutative side.
What is conventional in the latter, apart from the fact  taht
fields are multiplied with the ordinary product is that the transformation law for the
gauge field $A_\mu$ is governed by the ordinary covariant derivative,
\be
\delta_{\varepsilon} A_\mu(x) = A'_\mu(x) - A_\mu(x) =  D_\mu[A]
  \varepsilon(x)
\ee
Note that we are calling $\hat \varepsilon$, the infinitesimal gauge
transformation parameter in the noncommutative theory to distinguish
it from $\varepsilon$, its mapped relative in the ordinary theory.
Hence, the mapping
should include, appart from a connection between
$\hat A_\mu$ and $A_\mu$, one for connecting $\hat \varepsilon$ and $\varepsilon$.

The equivalence should hold at the
level of orbit space,
the physical configuration space of gauge
 theories.
 This means that if
two gauge fields $\hat A_\mu$ and $\hat A'_\mu$ belonging to the same orbit can be
connected by a noncommutative gauge transformation $\exp_*(i \hat \varepsilon)$,
then $A'_\mu$
and $A_\mu$, the corresponding mapped gauge fields will also be gauge
equivalent by an ordinary gauge transformetion $\exp(i \varepsilon)$.
An important
point is that  the mapping between $\hat \varepsilon$
and $\varepsilon$ necessarily depends on $A_\mu$. Indeed, were $\hat \varepsilon$
a function solely of $\varepsilon$, the ordinary and the noncommutative gauge
groups would be  identical. That this is not possible can be seen just by
considering the case of a  $U(1)$ gauge theory in which, through
a redefinition of the gauge parameter, one would be establishing an
isomorphism between non commutative $U_*(1)$ and commutative $U(1)$ gauge groups.

Then, the Seiberg-Witten mapping consists in finding
\ba
\hat A &=& \hat A [A;\theta] \nonumber \\
\hat \varepsilon &=& \hat \varepsilon [\varepsilon,
 A;\theta]
\ea
so that the   equivalence between orbits holds,
\be
\hat A [A] + \hat\delta_{\hat\varepsilon}\hat A [A] =
\hat A[A + \delta_\varepsilon A]
\label{sw}
\ee
Using the explicit form of gauge transformations and
expanding to first order in $\theta = \delta \theta$, the solution of
(\ref{sw}) reads
\begin{eqnarray}
&&\hat A_\mu[A] = A_\mu -\frac{1}{4}
 \delta\theta_{\rho \sigma}
 [A_\rho,\partial_\sigma A_\mu + F_{\sigma \mu}
 ]_+ + O(\delta\theta^2) \nonumber\\
&& \hat\varepsilon [\varepsilon,A] = \varepsilon + \frac{1}{4}
 \delta\theta_{\rho \sigma}[\partial_\rho,A_\sigma]_+ + O(\delta\theta^2)
 \label{solsw}
\end{eqnarray}
Concerning the field strength, the connection is given by
\be
\hat F_{\mu\nu}[A] = F_{\mu \nu} +\frac{1}{4}\delta\theta^{\alpha \beta}
\left(
2[F_{\mu \alpha},F_{\nu \beta}]_+ - [A_\alpha, D_\beta F_{\mu\nu}
+ \partial_\beta F_{\mu\nu}]_+
\right)
\label{swf}
\ee
One can interpret these equations as differential equations describing the passage from
$A_\mu^\theta$, the gauge field in a theory with parameter $\theta$ to
$A_\mu^{\theta + \delta \theta}$, the gauge field in a theory with parameter $\theta
+ \delta \theta$. Integrating of such equations one determines, to all orders in $\theta$,
how one passes from $L_{YM}[\hat A]$ the noncommutative version of Yang-Mills
Lagrangian, to $L[A,\theta]$ a complicated but commutative equivalent Lagrangian.

\subsection*{Fermion models in two dimensional space  and the W*Z*W
model}

It is well-known that for two-dimensional field theories, one can establish a connection,
called bosonization (fermionization),  that allows to connect a given fermion
(boson) model with certain bosonic (fermionic)
counterpart. This means that any physical quantity that can be computed for the fermionic
model can be alternatively computed for the bosonic model. Physical quantities are
constructed from correlation functions of products of currents and product of
energy-mementum tensor components (which in turn can be expressed in
terms of currents). The so called bosonization/fermionization recipe is nothing but  a
dictionary that tells how to connect the Lagrangian and the currents in  one
language to the Lagrangian and the currents in the other. Also, there
is a connection between coupling constants in the two models and, remarkably,
the weak coupling regime in one language corresponds to strong-coupling
regime in the other. This last property attracted, during  the 1970's and 1980's,
  a lot of people working in field theory and particle physics since bosonization
   provided
  a simplified laboratory where untractable non-perturbative calculations could be
  more easily handled perturbatively just by working in the equivbalen theory.

Non-Abelian bosonization (i.e., bosonization of a fermion model with non-Abelian
symmetry) was achieved in 1983 by Witten\cite{Wib} who showed that a a two-dimensional
theory of fermions $\psi$ in some representation of a non-Abelian group $G$ translates
into a bosonic theory where fields $a$ are group valued, $a \in G$. The connection
of the corresponding actions takes the form

~

\centerline{\small ~ bosonization/fermionization}

\vspace{- 0.3 cm}

\begin{center}
%WinTpicVersion2.16
\unitlength 0.1in
\begin{picture}(23.60,13.05)(22.90,-17.35)
\put(20.1000,-6.0000){\makebox(0,0)[lb]{$ \vphantom{{\rm
\,~ \;\;\; WZW}[a]} \int \!d^2x\, \bar\psi  i \!\!\not \! \partial
\psi$
}}%
\put(42.9000,-6.0000){\makebox(0,0)[lb]{${\rm ~ WZW}[a]$}}%
\put(22.7000,-12.6000){\makebox(0,0)[lb]{}}%
% VECTOR
\special{pn 8}%
\special{pa 2750 530}%
\special{pa 4230 530}%
\special{fp}%
\special{sh 1}%
\special{pa 4230 530}%
\special{pa 4163 510}%
\special{pa 4177 530}%
\special{pa 4163 550}%
\special{pa 4230 530}%
\special{fp}%
\special{sh 1}%
%%%%%%%%%%%%%%%%%%%%%%
\special{pa 2750 530}%
\special{pa 2817 510}%
\special{pa 2803 530}%
\special{pa 2817 550}%
\special{pa 2750 530}%
%%%%%%%%%%%%%%%%%%%%%%
\special{fp}%
\end{picture}%
\end{center}

\vspace{-2.9 cm}

\noindent{Figure 2: The bosonization/fermionization recipe connects the free
fermion action with a Wess-Zumino-Witten action}

~

\noindent Here, the Wess-Zumino-Witten action is given by
\ba
WZW[a] = \frac{1}{4\pi}{\rm tr^c}\!\!\int \!\!d^2x \left(
\partial_\mu a^{-1}\right) \left(\partial_\mu a\right)
 + \frac{i}{6\pi}\varepsilon_{ijk}{\rm tr^c}\!\!\int_B \!\!d^3y
a^{-1}(\partial_i a)  a^{-1}(\partial_j a) a^{-1}(\partial_k
a)\nonumber\\
\ea
where $B$ is a 3-dimensional manifold  which in
compactified Euclidean space can be identified with a ball with
boundary $S^2$. Index $i$ runs from 1 to 3.

The question we would like to pose is whether there exist a noncommutative version of
the bosonization recipe. That is, a connection of the form

~

~

\centerline{\small ~ bosonization/fermionization}

\vspace{- 0.3 cm}

\begin{center}
%WinTpicVersion2.16
\unitlength 0.1in
\begin{picture}(23.60,13.05)(22.90,-17.35)
\put(19.5000,-6.0000){\makebox(0,0)[lb]{$ \vphantom{{\rm
\,WZW}[\hat a]} \int \!d^2x\, \bar\psi * \! i \!\!\not \! \partial
\psi $
}}%
\put(42.9000,-6.0000){\makebox(0,0)[lb]{${\rm WZW}[\hat a]$}}%
\put(22.7000,-12.6000){\makebox(0,0)[lb]{}}%
% VECTOR
\special{pn 8}%
\special{pa 2750 530}%
\special{pa 4230 530}%
\special{fp}%
\special{sh 1}%
\special{pa 4230 530}%
\special{pa 4163 510}%
\special{pa 4177 530}%
\special{pa 4163 550}%
\special{pa 4230 530}%
\special{fp}%
\special{sh 1}%
%%%%%%%%%%%%%%%%%%%%%%
\special{pa 2750 530}%
\special{pa 2807 510}%
\special{pa 2803 530}%
\special{pa 2807 550}%
\special{pa 2750 530}%
%%%%%%%%%%%%%%%%%%%%%%
\special{fp}%
\end{picture}%
\end{center}

\vspace{-2.9 cm}

\noindent{Figure 3: A bosonization/fermionization recipe for noncommutative
 two dimensional models}

~

\noindent where the noncommutative version of
Wess-Zumino-Witten action would be
\ba
WZW[\hat a] &=& \frac{1}{4\pi}{\rm tr^c}\!\!\int \!\!d^2x \left(
\partial_\mu \hat a^{-1}\right) * \left(\partial_\mu \hat a\right)
+ \nonumber\\
&&  \frac{i}{6\pi}\varepsilon_{ijk}{\rm tr^c}\!\!\int_B \!\!d^3y
\hat a^{-1}*(\partial_i \hat a) * \hat a^{-1}*(\partial_j \hat a)
* \hat a^{-1}* (\partial_k
\hat a)
\label{w*z*w}
\ea
%\
where fields $\hat a$ are group valued, $\hat a \in G_*$. Now,  the noncommutative fermion action in the left hand side of
figure 3 being  quadratic, it should be equivalent to the ordinary fermion action
while the bosonic Wess-Zumino-Witten action  in the r.h.s.
contains cubic terms which cannot be trivially reduced
to  the commutative cubic terms of an ordinary WZW action. However,
going counter clockwise as indicated in figure 4, one could in principle pass
 from noncommutative to ordinary
WZW actions,

~

\begin{center}
%WinTpicVersion2.16
\unitlength 0.1in
\begin{picture}(23.60,13.05)(22.90,-17.35)
\put(20.1000,-6.0000){\makebox(0,0)[lb]{$ \vphantom{{\rm
\,WZW}[a]} \int \!d^2x\, \bar\psi\! *\! i \!\!\not \! \partial
\psi$
}}%
\put(21.4700,-18.2000){\makebox(0,0)[lb]{$\int\! d^2x\, \bar\psi
i\!\!\not
\! \partial \psi$}}%
\put(42.9000,-18.1000){\makebox(0,0)[lb]{${\rm \,WZW}[a]$}}%
\put(42.9000,-6.0000){\makebox(0,0)[lb]{${\rm WZW}[\hat a]$}}%
\put(22.7000,-12.6000){\makebox(0,0)[lb]{}}%
\put(46.5000,-12.7000){\makebox(0,0)[lb]{?}}%
% VECTOR
\special{pn 8}%
\special{pa 2950 530}%
\special{pa 4230 530}%
\special{fp}%
\special{sh 1}%
\special{pa 4230 530}%
\special{pa 4163 510}%
\special{pa 4177 530}%
\special{pa 4163 550}%
\special{pa 4230 530}%
\special{fp}%
\special{sh 1}%
%%%%%%%%%%%%%%%%%%%%%%
\special{pa 2950 530}%
\special{pa 3017 510}%
\special{pa 3003 530}%
\special{pa 3017 550}%
\special{pa 2950 530}%
%%%%%%%%%%%%%%%%%%%%%%
\special{fp}%
% VECTOR aqui
\special{pn 8}%
\special{pa 4520 1600}%
\special{pa 4520 700}%\special{pa 4520 740}%
\special{fp}%
\special{sh 1}%
\special{pa 4520 700}%\special{pa 4520 740}
\special{pa 4500 766}%\special{pa 4500 807}
\special{pa 4520 753}%\special{pa 4520 793
\special{pa 4540 766}%\special{pa 4540 807}
\special{pa 4520 700}%\special{pa 4520 740}
\special{fp}
%%%%%%%%%%%%%%%%%%%%%% flecha vert der abajo
\special{sh 1}%
\special{pa 4465 1620}%
\special{pa 4445 1553}%
\special{pa 4465 1570}%
\special{pa 4485 1553}%
\special{pa 4465 1620}%
%%%%%%%%%%%%%%%%%%%%%%
\special{fp}%
% CIRCLE
\special{pn 8}%
\special{ar 3590 1130 394 394  2.7109001 6.2831853}%
%\special{ar 3590 1130 394 394  2.7109001 6.2831853}%
%\special{ar 319 113 394 394  0.0000000 1.0957856}%
% SARROW
%\special{fp}%
\special{pn 8}%
%\special{pa 3748 1300}%
% \special{pa 3260 1250}%
%\special{pa 3230 1313}%
%\special{fp}%
\special{sh 1}%
\special{pa 3232 1300}%\special{pa 3230 1300}%
\special{pa 3200 1260}%\special{pa 3210 1250}%
\special{pa 3220 1270}%\special{pa 3230 1270}%
\special{pa 3240 1242}%\special{pa 3250 1250}%
\special{pa 3232 1300}%\special{pa 3230 1300}
\special{fp}%
% VECTOR 2 0 3 0
% 2 2950 2130 4230 2130
%
\special{pn 8}%
\special{pa 2950 1730}%
\special{pa 4230 1730}%
\special{fp}%
\special{sh 1}%
%%%%%%%%%%%%%%%%%%%%%%% flecha hori der abajo
\special{pa 4230 1730}%
\special{pa 4163 1710}%
\special{pa 4177 1730}%
\special{pa 4163 1750}%
\special{pa 4230 1730}%
\special{fp}%
\special{sh 1}%
%%%%%%%%%%%%%%%%%%%%%% flecha hori izq abaj
\special{pa 2950 1730}%
\special{pa 3017 1710}%
\special{pa 3003 1730}%
\special{pa 3017 1750}%
\special{pa 2950 1730}%
%%%%%%%%%%%%%%%%%%%%%%
\special{fp}%
% VECTOR 2 0 3 0
% 2 2610 2010 2600 1100
%
\special{pn 8}%
\special{pa 2600 1610}%
\special{pa 2600 700}%
\special{fp}%
\special{sh 1}%
%%%%%%%%%%%%%%%%%%%%%%%%% flecha vert izq arriba
\special{pa 2600 700}%
\special{pa 2570 766}%
\special{pa 2600 753}%
\special{pa 2621 766}%
\special{pa 2600 700}%
\special{fp}%
%%%%%%%%%%%%%%%%%%%%%% flecha vert izq abajo
\special{sh 1}%
\special{pa 2600 1620}%
\special{pa 2580 1553}%
\special{pa 2600 1570}%
\special{pa 2620 1553}%
\special{pa 2600 1620}%
%%%%%%%%%%%%%%%%%%%%%%
\special{fp}%
\end{picture}%
\end{center}

\vspace{0.3 cm}
\noindent{Figure 4: Connections between commutative and
noncommutative two dimensional models}

~

\noindent We shall clarify this issue using the path integral approach
to bosonization. This  requires as a first step  the knowledge of
the two dimensional (noncommutative) fermion determinant in
a background vector field. For simplicity, we shall   consider in detail
the $U_*(1)$ model
and then extend the treatment to the full $U_*(N)$ case.

\vspace{3.9 cm}

\noindent {\it {\bf The fermion determinant:
the $U_*(1)$ case}}

We now proceed to the exact calculation of the effective
action for noncommutative $U_*(1)$ fermions in the fundamental
representation as first presented in refs.[11]-[12], by
integrating the chiral anomaly. Indeed, taking profit that in 2
dimensions a gauge field $A_\mu$ can always be written in the
form\cite{Gamnpb}
\be \not \!\!A  = -\frac{1}{e}\left( i\!\!\not \!\partial
U[\phi,\eta] \right) * U^{-1}[\phi,\eta] \label{amu} \ee
with \be U[\phi,\eta] = \exp_*(\gamma_5 \phi + i \eta)  \, ,
\label{c} \ee one can relate the fermion determinant in a gauge
field background $A_\mu$ with that corresponding to $A_\mu = 0$
making a decoupling change of variables in the fermion fields.
For simpolicity, we consider here the case of fermions in
the fundamental representation (but calculations go the same in the other
two cases). The appropriate  change of
fermionic variables is \be \psi \to U[\phi,\eta]  * \psi \, ,
\hspace{2cm} \bar \psi \to \bar \psi * U^{-1}[\phi,\eta]
\label{cam} \ee
One gets\cite{Gaman}
\be {\det} ( \not \!\partial - ie\!\! \not \!\! A) = \det \not
\!\partial \exp\left( -2 \int_0^1 dt \frac{dJ^f[t\phi,A]}{dt}
\right)
\label{veinticinco} \ee
where $J^f[t\phi,A]$ is the Fujikawa Jacobian associated with a
transformation $U_t$ where $t$ is a parameter, $0 \leq t \leq1$, such that
\begin{eqnarray}
U_0 &=& 1 \nonumber
\\
U_1 &=& U[\phi,\eta]
\end{eqnarray}
Computation of the Jacobian $J^f[t\phi,A]$ for such finite
transformations $U_t$ can be done after  evaluation of the chiral  anomaly,
related to {\it infinitesimal} transformations. Indeed,
consider an infinitesimal local chiral transformation which in the
fundamental representation reads
\begin{eqnarray}
\delta_\epsilon^5 \psi &=& i \gamma_5\epsilon(x)* \psi(x)
\label{chi1}
\end{eqnarray}
The chiral anomaly ${\cal A} = {\cal A} ^a t^a$, associated with the non-conservatyion
of the chiral current $j^{a\mu}_5$
\be j^{a\mu}_5 = \bar \psi \gamma_5 t^a \psi \ee
\be
\partial_\mu j^{a\mu}_5 = {\cal A}^a[A] \, ,
\ee
 can be calculated from the Fujikawa Jacobian $J[\epsilon,A_\mu]$
associated with infinitesimal transformation (\ref{chi1}),
\be \log J[\epsilon,A_\mu] = -2 {\rm tr^c}\!\!\int \!\!d^2x  {\cal
A}[A]\epsilon(x) =\left. -2 {\rm tr^c}\!\! \int\!\! d^2x {\rm Tr}
\left(\gamma_5 \epsilon(x)\right)\right\vert_{reg} \label{canof}
\ee
Here  Tr means both a trace for Dirac matrices   and a functional trace
in the space on which the Dirac operator acts. With ${\rm tr}^c$ we
indicate a trace over the gauge group indices and with
{\it reg} we stress that some regularization prescription should be adopted
to render finite the Tr trace.  We shall adopt the heat-kernel
regularization, this meaning that
\be {\cal A}[A] = \lim_{M^2 \to \infty} {\rm Tr}\left( \gamma_5
\exp_*\left(\frac{\gamma^\mu \gamma^\nu D_\mu * D_\nu}{M^2}
\right) \right) \ee
The covariant derivative in the regulator has to be chosen  among
those defined  by eqs.(\ref{d1})-(\ref{d3}) according to the
representation one has chosen for the fermions.  Concerning the
fundamental representation,  the anomaly has been computed
following the standard Fujikawa procedure\cite{MS2}.
\be {\cal A}^f[A]  = \frac{e}{4\pi} \varepsilon^{\mu\nu}F_{\mu
\nu} \label{anof} \ee (We indicate with $f$ that the fundamental
representation has been considered).  Analogously, one obtains for
the anti-fundamental representation \be {\cal A}^{\bar f} [A]
=-\frac{e}{4\pi} \varepsilon^{\mu\nu}F_{\mu \nu} \label{anti} \ee

Now, writing $\epsilon = \phi dt$,  we can use the results given
through eqs.(\ref{canof})-(\ref{anof})  to get, for the Jacobian,
\be \log J^f[t\phi,A] = -\frac{e}{2\pi} {\rm tr^c} \int d^2x
\varepsilon_{\mu\nu}F^t_{\mu\nu} *\phi \label{put} \ee
where
\be F^t_{\mu\nu} = \partial_\mu A_\nu - \partial_\nu A_\mu
-ie\{A^t_\mu,A_\nu^t\} \ee \be \gamma^\mu A_\mu^t   =
-\frac{1}{e}\left( i\!\!\not \!\partial U_t \right) * U_t^{-1} \ee
This result can be put in a more suggestive way in the light cone
gauge where
\begin{eqnarray}
A_+ &=& 0 \nonumber\\ A_- &=&  g(x) *\partial_- g^{-1}(x)
\nonumber\\ g(x) &=& \exp_*(2\phi) \label{lcg}
\end{eqnarray}
Indeed, in this gauge one can see that (\ref{veinticinco}) becomes
\begin{eqnarray}
\log &&  \left(\frac{{\det} ( \not \!\partial -
ie\!\! \not \! A)}{{\det} \not \!\partial} \right)=\log
J^f[t\phi,A]  = - \frac{1}{8\pi}{\rm tr^c}\int d^2x \left(
\partial_\mu g^{-1}\right) *\left(\partial_\mu g\right) \nonumber\\
&&+  \frac{i}{12\pi}\varepsilon_{ijk}{\rm tr^c}\int_B d^3y
g^{-1}*(\partial_i g) * g^{-1}*(\partial_j g) *g^{-1}*(\partial_k
g)
\end{eqnarray}
Here we have written $d^3y = d^2xdt$  so that the integral in the
second line runs over a 3-dimensional manifold $B$ which in
compactified Euclidean space can be identified with a ball with
boundary $S^2$. Index $i$ runs from 1 to 3. The $*$ product on $B$
is the trivial extension of the noncommutative product defined in
the original two dimensional manifold, with the extra dimension
taken to be commutative. Concerning the anti-fundamental
representation, the calculation of the fermion determinant follows
identical steps.  Using the expression of the anomaly given by
(\ref{anti}), one computes the determinant which coincides with
that in the fundamental representation (remember that the anomaly
is proportional to the charge $e$ while the determinant to $e^2$).

For the Dirac operator  in the adjoint
representation, one starts from  the Dirac action which takes the form
\be S_f= \int d^2x  \bar \psi * i \gamma^\mu\left(\partial_{\mu}
\psi + i e \{A_{\mu}, \psi\}\right) \ee
Writing the field $A_{\mu}$ as in equation (\ref{amu}) (for
simplicity we will work in the gauge $\eta=0$) and making again a
change of the fermion variables to decouple the fermions from the
gauge fields
\begin{eqnarray}
\psi = e^{\gamma^5 \{\phi, \cdot\}} *\chi = \chi + \gamma^5
\{\phi, \chi\}+ \frac{1}{2} \{\phi,\{\phi,\chi\}\}
+\cdots\nonumber\\
{\bar \psi} = {\bar \chi} * e^{\{\cdot,\phi\} \gamma^5}  = {\bar
\chi}  +  \{{\bar \chi},\phi\} \gamma^5 + \frac{1}{2} \{\{{\bar
\chi},\phi\} ,\phi\} +\cdots \label{adcov}
\end{eqnarray}
one has
\be {\det} ( \not \!\partial + ie \gamma\cdot [A, \cdot]_*) = \det
\not \!\partial \exp\left( -2 \int_0^! dt
\frac{dJ^{ad}[t\phi,A]}{dt} \right) \ee
where $J^{ad}[t\phi,A]$ is the Fujikawa Jacobian associated with a
transformation for fermions in the adjoint (\ref{adcov}),
\begin{eqnarray}
\psi = e^{\gamma^5 [t \phi, \cdot]_*} *\chi \nonumber\\
{\bar \psi} = {\bar \chi} * e^{[\cdot,t \phi]_* \gamma^5}
\end{eqnarray}

Following the same procedure as before, one finds that
\be \frac{d}{dt} \log(J^{ad}) =\int d^2x\; {\cal A}(x)*\phi(x) \ee
where
\be {\cal A}(x) = 2 \lim_{M \to \infty} {\rm tr}\; \gamma^5\;
\int_k \exp_*({-i k\cdot x}) \exp_*\left({ \not \!\!
D^{2\;Ad}/M^2}\right) \exp_*({i k\cdot x}) \ee
After a straightforward computation, one can  prove the following
identity
\be \exp_*({-i k\cdot x}) \exp_*\left(\ds_{Ad}^{2}/M^2 \right)\;
\exp_*({i k\cdot x}) = \exp_*\left((\ds_{Ad} + i \ks + i e \cs)^2
/M^2\right) \ee
where
\begin{eqnarray}
c_{\mu}(x;k) &=& - 2 i \int_p\; \exp_* \left({-i p\wedge k}
\sin(p\wedge k)\right) A_{\mu}(p) \exp_*({i p\cdot x})\nonumber\\
&=& A_{\mu}(x-\theta\cdot k) - A_{\mu}(k)
\end{eqnarray}
Finally, expanding the exponent up to order $M^{-2}$, taking the
trace and integrating over $d^2 k$ we have
\be {\cal A}(x) = -\frac{e}{\pi} \lim_{M\to\infty}\; \int_p\;
\left((1-\exp_*\left({-M^2 \theta^2 p^2/4} \right)
\right)\epsilon_{\mu \nu} F_{\mu \nu}(p) \exp_*({i p\cdot x}) \ee
Notice that if we take the $\theta\to 0$ limit before taking the
$M\to \infty$ limit ${\cal A}$ vanishes and the Jacobian is $1$,
so we recover the standard (commutative) result which corresponds
to  a  trivial determinant for the trivial $U(1)$ covariant
derivative in the adjoint. Now, the limits $\theta\to 0$ and $M\to \infty$ do
not commute so that if one takes the $M^2 \to \infty$ limit at
fixed $\theta$ one has the $\theta$-independent result
\be \frac{d}{dt} \log J^{Ad}[t\phi,A] = -\frac{e}{\pi} \int d^2x
\varepsilon_{\mu\nu}F^t_{\mu\nu} *\phi \label{puti} \ee
which is twice the result of the fundamental representation. The
integral in $t$ is identical to the one of the fundamental
representation so we finally have
\begin{eqnarray}
\log &&\left(\frac{{\det}\left ( \not
\!\partial - ie\gamma_\mu\{A_\mu,~\} \right)}{{\det} \not
\!\partial} \right) = - \frac{1}{4\pi}{\rm tr^c}\int d^2x \left(
\partial_\mu g^{-1}\right) *\left(\partial_\mu g\right) \nonumber\\
&&+  \frac{i}{6\pi}\varepsilon_{ijk}{\rm tr^c}\int_B d^3y
g^{-1}*(\partial_i g) * g^{-1}*(\partial_j g) *g^{-1}*(\partial_k
g)
\end{eqnarray}
Comparing this result with that in the fundamental, we see that we
have proven the formula
\be \Gamma^{Adj} =2\; \Gamma^{Fund} \label{per66}\ee
This is reminiscent of the relation that holds
 when one  compares the anomaly and the fermion determinant
for commutative two-dimensional fermions in a  gauge field
background, for the fundamental and the adjoint representation of
$U(N)$. In this last case, there is a factor relating the results in
the adjoint and the fundamental which corresponds to the quadratic
Casimir $C(G)$ in the adjoint\cite{PW} (see
for example [57]  for a detailed derivation). Now, it was
observed [58] that diagrams in noncommutative
$U_*(1)$ gauge theories could be constructed in terms of those in
ordinary non-Abelian gauge theory with $C(G)=2$; this is precisely
what we have found in the present case.

We can stop at this point and recapitulate, using figure 4, the
results we have obtained: as indicated in the upper part of the figure,
we have connected the noncommutative fermion action with the noncommutative
version of the Wess-Zumino-Witten action which we suggestively write as $W*Z*W$.
The fact that the fermion action is quadratic allows us to descend to the ordinary
level from the left of the figure and bosonization connect this with an ordinary
$WZW$ action. The lacking link to close the loop in figure 4 should then
be a Seiberg-Witten like map except that in the present case it will not be related
to gauge symmetry but to the relevant symmetry in WZW theories. Moreover, it will
not connect two completely different Lagrangians but the two versions (commutative and
noncommutative) of the same Lagrangian.

~

\noindent {\it {\bf The Seiberg-Witten map for W*Z*W models}}

Before study the mapping between the non-commutative and standard
WZW theories, let us mention some properties of the Moyal
deformation in two dimensions.

Equation (\ref{moyalito}) can be re-written in term of holomorphic
and anti-holomor\-phic coordinates in the form:
\be \left.\phi(z,{\bar z})*\chi(z,{\bar z}) = \exp\left\{\theta
(\partial_z
\partial_{\bar w} -
\partial_{\bar z} \partial_w)\right\} \phi (z,{\bar z})
\chi(w,{\bar w})\right \vert_{w=z} \label{id1} \ee
This formula simplifies considerably in two particular cases. First,
when one of the functions is holomorphic (anti-holomorphic) and
the other is anti-holomorphic (holomorphic), the deformed product
reads as
\be \phi(z) * \chi(\bar z) = e^{\theta \partial_z \partial_{\bar
z}} \phi(z)\; \chi(\bar z) \;, \hspace{1cm} \phi(\bar z) * \chi(z)
= e^{-\theta \partial_z \partial_{\bar z}} \phi(\bar z)\; \chi(z)
\; \label{id2} \ee
and the deformation is produced by an overall operation over the
standard (commutative) product with no $w\to z$ limit necessary.

Second, when both functions are holomorphic (anti-holomorphic), the
star product coincides with the regular product
\be \phi(z)*\chi(z) = \phi(z) \chi(z)\; , \hspace{1cm} \phi({\bar
z})*\chi({\bar z}) = \phi({\bar z}) \chi({\bar z})\; . \label{id3}
\ee
That means that the holomorphic or anti-holomorphic sectors of a
two-dimensional field theory are unchanged by the deformation of
the product. For example, the holomorphic fermionic current in the
deformed theory takes the form,
\be
\hat {j}_z =\psi^{\dagger}_R\;* \psi_R\; . %\hspace{1cm}
\ee
And since $\psi_R$ has no ${\bar z}$ dependence on-shell, the
deformed current coincides with the standard one
\be
j_z =\psi^{\dagger}_R\; \psi_R\; . %\hspace{1cm}
\ee
Moreover, since the free actions are identical, any correlation
functions of currents in the standard and the $*$-deformed theory
will be identical.

This last discussion tell us that, since the WZW actions are the
generating actions of fermionic current correlation functions,
both actions (standard and non-commutative) are equivalent. It
remains to see if we can link both actions through a
Seiberg-Witten like mapping. Let us try that.

Consider the noncommutative  W*Z*W action, $W*Z*W[\hat g]$, with
$\hat g$ the gauge group element  when the
deformation parameter is $\theta$.
The action is invariant under
chiral holomorphic and anti-holomorphic transformations
\be \hat g \rightarrow \bar \Omega({\bar z})\hat g\; \Omega(z) \ee
so in analogy with the Seiberg-Witten mapping we will look for a
transformation that maps respectively holomorphic and
anti-holomorphic ``orbits" into ``orbits". Of course the analogy
breaks down at some point as this holomorphic and anti-holomorphic
``orbits" are not equivalence classes of physical configurations,
but just symmetries of the action. However we will see that such a
requirement is equivalent, in some sense, to the ``gauge orbits
preserving transformation condition" of Seiberg-Witten.

Thus, we will find a transformation that maps a group-valued field
$\hat g'$ defined in non-commutative space with deformation
parameter $\theta'$ to a group-valued field $\hat g$, with
deformation parameter $\theta$. We demand this transformation to
satisfy the condition
\be \bar \Omega'(\bar z)*' \hat g'*' \Omega'(z) \to \bar
\Omega(\bar z)* \hat g * \Omega(z) \label{chiraltraf} \ee
where the primed quantities are defined in a
$\theta'$-non-commutative space and the non primed quantities
defined in a $\theta$-non-commutative space. In particular this
mapping will preserve the equations of motion:
\be \hat g' = \alpha' *' \beta' \to \hat g=\alpha * \beta \ee

The simplest way to achieve this, by examining equation
(\ref{id2}) is defining
\be \hat g[\theta] = e^{-\theta\partial_z\partial_{\bar z}}\; g[0]
\ee
or, infinitesimally
\be \frac{d \hat g}{d\theta}  = - \partial_z\partial_{\bar z}\;
\hat g \label{traf1} \ee
However, the corresponding transformation for ${\hat g}^{-1}$ is
more cumbersome
\be \frac{d {\hat g}^{-1}}{d\theta}  = \partial_z\partial_{\bar
z}\; {\hat g}^{-1} + 2 \partial_{\bar z}({\hat g}^{-1}*\partial_z
\hat g)*{\hat g}^{-1} \label{traf2} \ee
So let us consider a more symmetric transformation, that coincides
on-shell, with (\ref{traf1}) and (\ref{traf2}). Consider thus
\begin{eqnarray}
\frac{d \hat g}{d\theta} &=& \hat g*\partial_{\bar z}\;
{\hat g}^{-1}*\partial_z \hat g \nonumber \\
\frac{d {\hat g}^{-1}}{d\theta} &=& -\partial_z {\hat
g}^{-1}*\partial_{\bar z} \hat g *{\hat g}^{-1} \label{trafII}
\end{eqnarray}
These equations satisfy the condition (\ref{chiraltraf}) for
functions $\Omega(z)$ and $\bar \Omega(\bar z)$ independent of
$\theta$. Indeed we have, for example
\begin{eqnarray}
\frac{d (\hat g * \Omega(z))}{d \theta}  &=& \frac{d \hat g}{d
\theta}* \Omega(z) - \partial_{\bar z}\hat g * \partial_z
\Omega(z) \nonumber\\
&=& (\hat g*\Omega(z))*\partial_{\bar z}\;(\hat g *
\Omega(z))^{-1} * \partial_z (\hat g* \Omega(z)) \label{inv1}
\end{eqnarray}
and a similar equation for the anti-holomorphic transformation.

The next step is to see how does the WZW action transforms under
this mapping. First consider the variation of the following
object:
\be \omega = {\hat g}^{-1}*\delta \hat g \label{omega} \ee
where $\delta$ is any variation that does not acts on $\theta$.

After a straightforward computation we find that
\be \frac{d \omega}{d\theta}  = - \partial_{\bar z} \omega * j_z -
j_z
* \partial_{\bar z} \omega
\label{trafom} \ee
where
\be j = {\hat g}^{-1}*\partial_z \hat g \label{j1} \ee
is the holomorphic current. In particular we have
\be \frac{d j_z}{d\theta}  = - \partial_{\bar z} j_z * j_z - j_z
* \partial_{\bar z} j_z = -\partial_{\bar z} (j_z^2)
\label{trafj} \ee
Similarly we can find the variations for
\be \bar \omega = \delta \hat g * {\hat g}^{-1} \label{bomega} \ee
and we get
\begin{eqnarray}
\frac{d \bar \omega}{d\theta}  &=&  \partial_{z} \omega * j_{\bar
z} + j_{\bar z}* \partial_{z} \bar \omega \nonumber\\
\frac{d j_{\bar z}}{d\theta}  &=& \partial_{z} j_{\bar z} *
j_{\bar z} + j_{\bar z}* \partial_{z} j_{\bar z} = \partial_z
(j_{\bar z}^2) \label{trafbar}
\end{eqnarray}
where
\be j_{\bar z} = \partial_{\bar z} \hat g * {\hat g}^{-1} .
\label{barj} \ee
Note that on-shell, both $j_z$ and $j_{\bar z}$ are
$\theta$-independent, that is the non-commu\-ta\-tive currents
coincide with the standard ones. This result is expected since the
same happens for their fermionic counterparts.

Now, instead of studying how does the $\theta$-map acts on the WZW
action, it is easy to see how does the mapping acts on the {\sl
variation} of the WZW action with respect to the fields. In fact,
we have
\be \delta W[\hat g] = \frac{k}{\pi}\int d^2x\; {\rm
tr}\left(\partial_{\bar z}j_z\; \omega\right) \label{varwzw} \ee
where $j_z$ and $\omega$ are the quantities defined in
eqs.(\ref{omega}) and (\ref{j1}) and there is no $*$-product
between them in eq.(\ref{varwzw}) in virtue of the quadratic
nature of the expression.

Thus, a simple computation shows that
\be \frac{d\; \delta W[\hat g]}{d\theta} = 0 \label{st} \ee
and we  have a remarkable result: the transformation
(\ref{trafII}), integrated between $0$ and $\theta$ maps the
standard commutative WZW action into the noncommutative WZW
action. That is, we have found a transformation mapping orbits
into orbits such that it keeps the form of the action unchanged
provided one simply performs a $\theta$-deformation. This should
be contrasted with the 4 dimensional noncommutative Yang-Mills
case for which a mapping respecting gauge orbits can be found (the
Seiberg-Witten mapping) but the resulting commutative action is
not the standard Yang-Mills one. However, one can see that the
mapping (\ref{trafII}) is in fact a kind of Seiberg-Witten change
of variables.

Indeed, if we consider the WZW action as the effective action of a
theory of Dirac fermions coupled to gauge fields, as we did in
previous sections, instead of an independent model, we can relate
the group valued field $g$ to gauge potentials. As we showed in
eq.(\ref{lcg}), this relation acquires a very simple form in the
light-cone gauge $A_+=0$ where
\be A_- =  \hat g(x) *\partial_- {\hat g}^{-1}(x)\; . \ee
But notice that in this gauge, $A_-$ coincides with $j_{\bar z}$
(eq.(\ref{barj}), so we have from equation (\ref{trafbar})
\be \delta A_- = \delta \theta \left(\partial_{z} A_- * A_- + A_-*
\partial_{z} A_- \right)
\ee
which is precisely the Seiberg-Witten mapping in the gauge
$A_+=0$.

~

\centerline{\it The loop in figure 4 is now closed}

\subsection*{C*S theory in d=3 dimensions}
The W*Z*W action was obtained by computing the fermion determinant of the
Direc operator for $d=2$ noncommutative fermions
coupled to a gauge-field background. Thus,
it is natural to compute the same determinant, but in $d=3$ spacetime.
This was done by Grandi and Silva\cite{GrSi} and we shall
briefly describe the results they have found.
For simplicity, we take massive fermions so that a $\partial/m$
expansion can be easily used to compute the parity odd part of the
determinant.

Coupling fermions to a gauge
field $A_\mu$ in the Lie Algebra of $U_*(N)$, the action reads
\be
 S(\bar\psi,\psi,  A;m) \;=\;  \int d^3 x \; {\bar { \psi}}({x})* ( i\not
 \!\! D[A] -m ) ({x}) \; .
 \label{fund}
\ee
and again we have three possibilities for the Dirac operator
\be
 \psi(x)
 \to\left\{
 \begin{array}{ll}
  g(x)*\psi(x)&${\rm fundamental representation} $``f"$ $\\
 \psi(x)*  g^{-1}(x)&${\rm anti-fundamental representation}
 $``\bar f"$ $\\
 g(x)*(x)*g^{-1}(x)&${\rm adjoint representation}
 $``ad"$ $
 \end{array}
 \right.
\ee
Accordingly, the covariant derivative acting on $\psi$ takes the form
\be
  D_\mu [A]  =
 \left\{
 \begin{array}{ll}
 \partial_\mu  + i e \,{ A}_\mu\!* &
 $ $``f" $ $
 \\
 \partial_\mu  - i e\, *  A_\mu &
 $ $``\bar f"$ $
 \\
 \partial_\mu  + i e\,[ A_\mu,~] &
 $ $``ad"$ $
 \end{array}
 \right.~.
 \label{defi}
\ee
The effective action $\Gamma(\hat A;m)$ is defined as
\be
e^{i\Gamma(A; m)} =
Z(A; m)
=
\int{  D}{\psi}{ D}{\bar{\psi}}
\,e^{iS(\bar \psi, \psi,A; m)}
\label{Z}
\ee
Calculation of the effective action for fermions in the
fundamental and the anti-fundamental representations gives the
same answer. Consider the case of the fundamental. As in the original
calculation of induced CS actions\cite{redlich},
one can obtain the contribution to $\Gamma_{odd}(\hat A; m)$  from
the vacuum polarization and the triangle graphs
\ba
 \label{sss}
 i \Gamma_{odd} [\hat A;m] &=& \left. \left(
 \frac{1}{2} {\rm tr^c} \int \frac{ d^3 p}{(2\pi)^3}  {\hat A}_\mu
 (p) \,\Pi^{\mu\nu}(p;m)\, {\hat A}_\nu (-p)  + \right. \right. \n
 && \!\!\!\!\!\!\!\!\!\! \left. \left. +\frac{1}{3}{\rm tr^c}
 \int \frac{ d^3 p}{(2\pi)^3}\frac{ d^3
 q}{(2\pi)^3}\Gamma^{\mu\nu\rho}(p,q;m)\,{\hat A}_\mu(p) {\hat
 A}_\nu(q) {\hat A}_\rho (-p-q) \right) \right\vert_{odd}
 \n
 \label{unoo}
\ea
As before,  ${\rm tr^c}$  represents the trace over the $U(N)$ algebra generators,
and
\be
\label{111}
 \Pi^{\mu \nu}(p;m) = -e^2
\int \frac { d^3 k}{(2\pi)^3} {\rm tr} \left[ \gamma^\mu
\frac{\slash\!\!\!k -m}{k^2 -m^2}\gamma^\nu \frac{\slash\!\!\!k
+\slash\!\!\!\!p - m} {(k+p)^2 -m^2} \right] \ee
\ba
 \label{222}
 \Gamma^{\mu\nu\rho}(p,q;m) \!\!\!&=&\!\!\!
 e^3\,\exp(- \frac{i}{2} p_\lambda \theta^{\lambda\delta} q_\delta
 ) \int \!\frac{ d^3 k}{(2\pi)^3} {\rm tr}\! \left[\gamma^\mu
 \frac{(\slash\!\!\!k - m)}{k^2 -m^2} \gamma^\nu \frac{
 (\slash\!\!\!k-\slash\!\!\!q -m)}{(k-q)^2 -m^2}\times \right.
 \n
 && \;\;\;\; \left. \times \gamma^\rho
 \frac{(\slash\!\!\!k +\slash\!\!\!p -m)}{(k+p)^2-m^2} \right]
\ea
There are
no nonplanar contributions to the parity odd sector of the
effective action\cite{chu}. The only modification arising from
noncommuativity is the $\theta$-dependent phase factor in
$\Gamma^{\mu\nu\rho}$, associated to external legs in the cubic
term, which is nothing but the star product in configuration
space. The result for $\Gamma_{odd}(\hat A;m)$ is analogous to the
commutative one except that the star $*$-product replaces the
ordinary product.

Regularization of the divergent integrals (\ref{111}) and
(\ref{222}) can be achieved by introducing in the original action
(\ref{fund}), bosonic-spinor Pauli-Villars fields with mass $M$.
These fields  give rise to additional diagrams, identical to those
of eq.(\ref{unoo}), except that the regulating mass $M$ appears in
place of the physical mass $m$. Since we are interested in the
parity violating part of the effective action, we keep only the
parity-odd terms in (\ref{111}) and (\ref{222}) (and in the
corresponding regulator field graphs). To leading order in
$\partial/m$, the gauge-invariant parity violating part of the
effective action is, for the fundamental representation, given by
\ba
 \Gamma_{odd}^f(\hat A,m)
 &=& \frac{1}{2}
 \left( \frac{m}{|m|}+ \frac{M}{|M|} \right) {\hat S}_{CS}(\hat A)
 +O(\partial^2/m^2)
 \n
 &=& \pm  {\hat S}_{CS}(\hat A)
 +O(\partial^2/m^2)
 \label{mM}
\ea
with
\be
{\hat S}_{CS}(\hat A) = \frac{e^2}{4\pi}\int d^3x\, \epsilon^{\mu\nu\rho} {\rm
tr} \left( \hat A_\mu *
\partial_\nu \hat A_\rho\right.\nonumber\\
\left.+  \frac{ 2ie}{ 3} \hat A_\mu* \hat A_\nu* \hat A_\rho
\right)
\label{ec}
 \ee
As it is well known, the relative sign of the fermion and
regulator contributions depends on the choice of the Pauli-Villars
regulating Lagrangian (of course the divergent parts should cancel
out independently of this choice). In the first line of (\ref{mM})
we have made a choice such that the two contributions add to give
the known Chern-Simons result of the second line. Note that even
in the Abelian case, the Chern-Simons action contains a cubic term
(analogous to that arising in the ordinary non-Abelian case).

Calculations for fermions in the anti-fundamental representation follows the same
steps. There is just a change of sign $e \to -e$ on each vertex,
compensated by a change in the momenta dependence of propagators
due to the different ordering of fields in the $f$ and $\bar f$
covariant derivatives (see (\ref{defi})). The result
then coincides with the fundamental one.

Concerning the adjoint representation, calculations are a little more lengthy
and I shall just give the answer\cite{GrSi} to
leading order in $\partial/m$,
\be
\Gamma_{odd}^{ad}(\hat A,m) =\pm
 \hat S_{CS}(\hat A) +  O(\partial^2/m^2)~.
 \label{com}
\ee
As before, the result is gauge invariant even under large gauge
transformations.

 It should be stressed that (\ref{com}) gives a non-trivial effective
action even in the   $\theta \to 0$ limit, in which fermions in
the adjoint decouple from the gauge field. As already explained in
the  two
dimensional case, this is due to
the fact that this limit does not commute with that of the
regulator $M \to \infty$.

Now, an enhancement of  Fig.4 can be envisaged,

\vspace{-2 cm}

\begin{center}
%WinTpicVersion2.16
\unitlength 0.1in
{\hspace{-4.7 cm} \begin{picture}(23.60,23.05)(22.90,-17.35)
\put(20.1000,-6.0000){\makebox(0,0)[lb]{$ \vphantom{{\rm
\,WZW}[a]} \int \!d^2x\, \bar\psi\! *\! i \!\!\not \! \partial
\psi$
}}%
\put(21.4700,-18.2000){\makebox(0,0)[lb]{$\int\! d^2x\, \bar\psi
i\!\!\not
\! \partial \psi$}}%
\put(42.9000,-18.1000){\makebox(0,0)[lb]{${\rm \,WZW}[a]$}}%
\put(42.9000,-6.0000){\makebox(0,0)[lb]{${\rm WZW}[\hat a]$}}%
\put(62.9000,-18.1000){\makebox(0,0)[lb]{${\rm \,CS}[a]$}}%
\put(62.9000,-6.0000){\makebox(0,0)[lb]{${\rm C*S}[\hat a]$}}%
\put(22.7000,-12.6000){\makebox(0,0)[lb]{}}%
\put(66.5000,-12.7000){\makebox(0,0)[lb]{?}}%
% VECTOR
\special{pn 8}%
\special{pa 2950 530}%
\special{pa 4230 530}%
\special{fp}%
\special{sh 1}%
\special{pa 4230 530}%
\special{pa 4163 510}%
\special{pa 4177 530}%
\special{pa 4163 550}%
\special{pa 4230 530}%
\special{fp}%
\special{sh 1}%
%%%%%%%%%%%%%%%%%%%%%%
\special{pa 2950 530}%
\special{pa 3017 510}%
\special{pa 3003 530}%
\special{pa 3017 550}%
\special{pa 2950 530}%
%%%%%%%%%%%%%%%%%%%%%%
\special{fp}%
% VECTOR aqui
\special{pn 8}%
\special{pa 4520 1600}%
\special{pa 4520 700}%\special{pa 4520 740}%
\special{fp}%
\special{sh 1}%
\special{pa 4520 700}%\special{pa 4520 740}
\special{pa 4500 766}%\special{pa 4500 807}
\special{pa 4520 753}%\special{pa 4520 793
\special{pa 4540 766}%\special{pa 4540 807}
\special{pa 4520 700}%\special{pa 4520 740}
\special{fp}
% VECTOR aqui
\special{pn 8}%
\special{pa 6420 1600}%
\special{pa 6420 700}%\special{pa 4520 740}%
\special{fp}%
\special{sh 1}%
\special{pa 6420 700}%\special{pa 4520 740}
\special{pa 6400 766}%\special{pa 4500 807}
\special{pa 6420 753}%\special{pa 4520 793
\special{pa 6440 766}%\special{pa 4540 807}
\special{pa 6420 700}%\special{pa 4520 740}
\special{fp}
%%%%%%%%%%%%%%%%%%%%%% flecha vert der abajo
\special{sh 1}%
\special{pa 6375 1620}%
\special{pa 6355 1553}%
\special{pa 6375 1570}%
\special{pa 6395 1553}%
\special{pa 6375 1620}%
%%%%%%%%%%%%%%%%%%%%%%
\special{fp}%
%%%%%%%%%%%%%%%%%%%%%% flecha vert der abajo
\special{sh 1}%
\special{pa 4430 1620}%
\special{pa 4410 1553}%
\special{pa 4430 1570}%
\special{pa 4450 1553}%
\special{pa 4430 1620}%
%%%%%%%%%%%%%%%%%%%%%%
\special{fp}%
% CIRCLE
\special{pn 8}%
\special{ar 3590 1130 394 394  2.7109001 6.2831853}%
%\special{ar 3590 1130 394 394  2.7109001 6.2831853}%
%\special{ar 319 113 394 394  0.0000000 1.0957856}%
% SARROW
%\special{fp}%
\special{pn 8}%
%\special{pa 3748 1300}%
% \special{pa 3260 1250}%
%\special{pa 3230 1313}%
%\special{fp}%
\special{sh 1}%
\special{pa 3232 1300}%\special{pa 3230 1300}%
\special{pa 3200 1260}%\special{pa 3210 1250}%
\special{pa 3220 1270}%\special{pa 3230 1270}%
\special{pa 3240 1242}%\special{pa 3250 1250}%
\special{pa 3232 1300}%\special{pa 3230 1300}
%%%%%%%%%%%%%%%%%%%%%%%%%%%%%%%%%%%%%%%%%%%%%%%%%%%
%%%%%%%%%%%%%%%%%%%%%%%%%%%%%%%%%%%%%%%%%%%%%%%%%%%
\special{fp}%
% CIRCLE
\special{pn 8}%
\special{ar 5590 1130 394 394  2.7109001 6.2831853}%
% SARROW
%\special{fp}%
\special{pn 8}%
\special{pa 5232 1300}%\special{pa 3230 1300}%
\special{pa 5200 1260}%\special{pa 3210 1250}%
\special{pa 5220 1270}%\special{pa 3230 1270}%
\special{pa 5240 1242}%\special{pa 3250 1250}%
\special{pa 5232 1300}%\special{pa 3230 1300}
%%%%%%%%%%%%%%%%%%%%%%%%%%%%%%%%%%%%%%%%%%%%%%%%%%
%%%%%%%%%%%%%%%%%%%%%%%%%%%%%%%%%%%%%%%%%%%%%%%%%%%
\special{fp}%
% VECTOR 2 0 3 0
% 2 2950 2130 4230 2130
%
\special{pn 8}%
\special{pa 2850 1730}%
\special{pa 4180 1730}%
\special{fp}%
% VECTOR 2 0 3 0
\special{pn 8}%
\special{pa 4750 530}%
\special{pa 6130 530}%
\special{fp}%
% VECTOR 2 0 3 0
\special{pn 8}%
\special{pa 4750 1730}%
\special{pa 6130 1730}%
\special{fp}%
\special{sh 1}%
%%%%%%%%%%%%%%%%%%%%%%% flecha hori extrema der abajo
\special{pa 6180 530}%
\special{pa 6113 510}%
\special{pa 6127 530}%
\special{pa 6113 550}%
\special{pa 6180 530}%
\special{fp}%
\special{sh 1}%
%%%%%%%%%%%%%%%%%%%%%% flecha hori extrema izq abaj
\special{pa 4750 530}%
\special{pa 4817 510}%
\special{pa 4803 530}%
\special{pa 4817 550}%
\special{pa 4750 530}%
%%%%%%%%%%%%%%%%%%%%%%
\special{fp}%
\special{sh 1}%
%%%%%%%%%%%%%%%%%%%%%%% flecha hori extrema der abajo
\special{pa 6180 1730}%
\special{pa 6113 1710}%
\special{pa 6127 1730}%
\special{pa 6113 1750}%
\special{pa 6180 1730}%
\special{fp}%
\special{sh 1}%
%%%%%%%%%%%%%%%%%%%%%% flecha hori extrema izq abaj
\special{pa 4750 1730}%
\special{pa 4817 1710}%
\special{pa 4803 1730}%
\special{pa 4817 1750}%
\special{pa 4750 1730}%
%%%%%%%%%%%%%%%%%%%%%%
\special{fp}%
%%%%%%%%%%%%%%%%%%%%%%% flecha hori der abajo
\special{pa 4180 1730}%
\special{pa 4113 1710}%
\special{pa 4127 1730}%
\special{pa 4113 1750}%
\special{pa 4180 1730}%
\special{fp}%
\special{sh 1}%
%%%%%%%%%%%%%%%%%%%%%% flecha hori izq abaj
\special{pa 2850 1730}%
\special{pa 2917 1710}%
\special{pa 2903 1730}%
\special{pa 2917 1750}%
\special{pa 2850 1730}%
%%%%%%%%%%%%%%%%%%%%%%
\special{fp}%
% VECTOR 2 0 3 0
% 2 2610 2010 2600 1100
%
\special{pn 8}%
\special{pa 2600 1610}%
\special{pa 2600 700}%
\special{fp}%
\special{sh 1}%
%%%%%%%%%%%%%%%%%%%%%%%%% flecha vert izq arriba
\special{pa 2600 700}%
\special{pa 2570 766}%
\special{pa 2600 753}%
\special{pa 2621 766}%
\special{pa 2600 700}%
\special{fp}%
%%%%%%%%%%%%%%%%%%%%%% flecha vert izq abajo
\special{sh 1}%
\special{pa 2600 1620}%
\special{pa 2580 1553}%
\special{pa 2600 1570}%
\special{pa 2620 1553}%
\special{pa 2600 1620}%
%%%%%%%%%%%%%%%%%%%%%%
\special{fp}%
\end{picture}%
}
\end{center}

\vspace{0.3 cm}

\noindent{Figure 5: The connections now imply three dimensional models}

~

The  new  connection in the right lower part of the figure
is well-established: one can relate the 3-dimensional
 Chern-Simons
action with the 2-dimensional WZW model using different
approaches\cite{Wics,Alr}. It is this well-known connection that suggests
the upper one, its noncommutative version. Now, if such
a connection holds one should expect that,
as it happens for the Wess-Zumino-Witten
model,  a Seiberg-Witten map will allow to pass (the arrow with the interrogation
mark) from  noncommutative
to commutative Chern-Simons theories.

Let us the start by connecting $C*S$ with $W*Z*W$ actions.
Consider the expression
\be
 \hat S_{CS}[\hat A_0,\hat A_i]=\frac { e^2} {4\pi}{\rm Tr} \int_{{\cal M}}
 d^3x\, \epsilon_{ij}\left({\hat A}_0* {\hat F}_{ij}+ \dot {\hat
 A}_i *{\hat A}_j\right)~,
 \label{wzw1}
\ee
which differs from the  CS action
(\ref{ec}) by a surface term. Of course,  when ${\cal M}$ has no
boundary, such surface term is irrelevant.  However, in what
follows we choose as manifold ${\cal M} = \Sigma \times R$ with
$\Sigma$ a two-dimensional manifold. We shall take eq.(\ref{wzw1})
as the starting point for quantization of the $2+1$ theory and
follow the steps described in [69]-[71] in the
original derivation of the (ordinary space) connection and in [62] for
the noncommutative extension.

Expression  (\ref{wzw1}) can be rewritten as
\be
 \hat S_{CS}[\hat A_0,\hat A_i]= \frac { e^2} {4\pi}{\rm Tr} \int_{{\cal
 M}} d^3x\, \epsilon_{ij}\left({\hat A}_0 {\hat F}_{ij} + \dot {\hat
 A}_i *{\hat A}_j\right) + \frac { e^2} {4\pi}{\rm Tr} \int_{\partial {\cal
 M}} dS_{\mu}\Lambda^{\mu}
 \label{wzw2}
\ee
with
\be
 \Lambda^\mu = \epsilon_{ij}\sum_{n=1}^{\infty}\frac 1 {n!}
 \left(\frac i 2\right)^n \theta^{\mu
 \nu_1}\theta^{\mu_2\nu_2}...\theta^{\mu_n\nu_n}
 \partial_{\mu_2}...\partial_{\mu_n}{\hat A}_0\,\partial_{\nu_1}
 \partial_{\nu_2}...\partial_{\nu_n}{\hat F}_{ij}
\ee
 Using action (\ref{wzw2}),  the partition function for
 the noncommutative C*S theory takes the form
\ba
 Z &=& \int {\cal D}\hat A_i {\cal D}\hat A_0
 \exp\left( \frac {i\kappa e^2} {4\pi}{\rm Tr}\int_{{\cal
 M}} d^3x\, \epsilon_{ij}\left({\hat A}_0 {\hat F}_{ij}+ \dot {\hat
 A}_i *{\hat A}_j\right)\right. \n
 &&\left.\;\;\;\;+\frac {i\kappa e^2} {4\pi}{\rm Tr}\int_{\partial{\cal
 M}}dS_{\mu}\Lambda^{\mu}\right)
 \label{z3}
\ea
where $\kappa$ is an integer. For interior points of ${\cal M}$,
$A_0$ acts as a Lagrange multiplier enforcing flatness of the
spatial components of the connection
\be
\hat F_{ij}(x) = 0 \;\;\;\;\;\;\;\;\; \forall x\in {\cal
M}-\partial{\cal M}
\label{flat}
\ee
By continuity, $\hat F_{ij}$ must also vanishes on the boundary.
The partition function takes then form
\be
 Z= \int {\cal D}\hat A_i \delta(\epsilon_{ij}\hat F_{ij})
 \exp\left(\frac {i\kappa e^2} {4\pi}{\rm Tr}\int_{\cal M}
 d^3x\,\epsilon_{ij}\dot{\hat A}_i* \hat A_j\right)
 \label{z22}
\ee

Let us  discuss the case where $\Sigma$ is the
disk.  In that case,  the solution of the flatness condition (\ref{flat}) is
$
\hat A_i = - \frac i e \,\hat g^{-1} *
\partial_i \hat g
$,
and one has, reinserting it in (\ref{z22}),
\be
Z = \int {\cal D}\hat g \exp\left(i\kappa\hat S_{CWZW}[\hat
g]\right) \label{zz3} \ee
where $\hat S_{CWZW}[\hat g]$ is the noncommutative, chiral WZW
action
%first discussed in \cite{MS},
%
\ba
 \hat S_{CWZW}[\hat g] &=& -\frac 1 {4\pi} {\rm Tr}\int_{\partial {\cal M}}
 d^2x (\hat g^{-1}*\partial_t \hat g)*(\hat
 g^{-1}*\partial_\varphi \hat g)
 \nonumber\\
 & &-\frac 1 {4\pi }{\rm Tr} \int_{{\cal M}} d^3x
 \,\epsilon_{ij}(\hat g^{-1} * \partial_i g)* ( \hat g^{-1} *
 \partial_t g)*( \hat g^{-1} * \partial_j g)
 \n
\ea
here $\varphi$ is a tangential coordinate which parametrize the
boundary of ${\cal M}_2$.

We have then seen that the upper W*Z*W$ \leftrightarrow $C*S connection in Fig.5
 has been established.
One should then expect that a Seiberg-Witten map  will provide the $\updownarrow$
connection, closing the second loop in the figure.
Let us write a generic Seiberg-Witten map (see eqs.(\ref{solsw})-(\ref{swf}) in the form
\ba
\delta\hat A^\mu &=& \delta \theta^{\rho\sigma}
\frac \partial {\partial\theta^{\rho\sigma} } \hat
A_{\mu}(\theta)=
-\frac 1 4 \delta \theta^{\rho\sigma}
\left[
\hat A_\rho ,
\partial_\sigma \hat A_\mu+\hat F_{\sigma\mu}
\right]_+
\n
\delta\hat F_{\mu\nu}(\theta)&=&
\delta \theta^{\rho\sigma}
\frac \partial {\partial\theta^{\rho\sigma} } \hat
F_{\mu\nu}(\theta)
=
\frac 1 4 \delta\theta^{\rho\sigma}
\left(
2
\left[
 \hat F_{\mu\rho},\hat F_{\nu\sigma}
\right]_+
-
\left[
\hat A_\rho,\hat D_\sigma\hat F_{\mu\nu}+\partial_\sigma\hat F_{\mu\nu}
\right]_+
\right)\nonumber\\
\label{swt}
\ea
Starting from (\ref{ec}) with ${\cal M}$ either $R^3$ or $\Sigma\times R$
with $\Sigma$ a   manifold without boundary
 the
noncommutative  Chern-Simons action  can be written in the form
\be
 \hat S_{CS}(\hat A)=
 \frac{e^2}{4\pi}\int_{{\cal M}} d^3x\, \epsilon^{\mu\nu\rho}
 \left(
 \hat A_\mu * \partial_\nu\hat A_\rho
 +
 \frac {2ie}{3}
 \hat A_\mu* \hat A_\nu* \hat A_\rho
 \right)
 \label{15}
\ee
or
\ba
 \hat S_{CS}(\hat A) &=&\frac{e^2}{4\pi}\int_{{\cal M} }d^3x \,
 \epsilon_{ij} \left( {\hat  A}_0 {\hat F}_{ij} + \dot {\hat A}_i
 \hat  A_j \right)~.
 \label{F2}
\ea
Differentiate this expression with respect to
$\theta_{\mu\nu}$ one gets
\ba
 \frac{\partial {\hat S_{CS}[\hat A,\theta]}}{\partial\theta_{\mu\nu}}
 &=&\frac{e^2}{4\pi}\int_{{\cal M} } d^3x \,
 \epsilon_{ij}
 \frac \partial{\partial
 \theta_{\mu\nu}} \left( \hat A_0 \hat  F_{ij} +\dot A_i  A_j
 \right)\n
 &=&\frac{e^2}{4\pi}\int_{{\cal M} }d^3x \,
 \epsilon_{ij}
 \left( \frac {\partial
 \hat A_0}{\partial \theta_{\mu\nu}} \hat  F_{ij} + \hat A_0
 \frac{\partial \hat  F_{ij}}{\partial \theta_{\mu\nu}} + 2 \frac
 {\partial A_j}{\partial \theta_{\mu\nu}}\dot A_i \right)
\ea
Then, using (\ref{swt})   to rewrite the
$\theta$-derivatives and keeping just the terms which are
antisymmetric with respect to the indices $\mu,\nu$ and $i,j$, we
get
\ba
 \frac{\partial {\hat S_{CS}[\hat A,\theta]}}{\partial\theta_{\mu\nu}}
 = 0
 \label{a}
\ea
this meaning that \be S_{CS}[A,\theta] = S_{CS}[A,0] \ee Here
$S_{CS}[A,0]$ is just the ordinary (commutative) CS action. It is
interesting to note that in the $U_*(1)$ case the SW map cancels
out the cubic term which is present in $\hat S_{CS}(\hat A)$. We
have then seen  how  the Seiberg-Witten transformation (\ref{swt})
allows to pass from noncommutative to ordinary Chern-Simons
action.

~

\centerline{\it The second loop is closed.}

~

Let us end this lecture by noting  that the exact parity-breaking
part of the effective action for 2+1 QED at {\it finite
temperature} can be easily computed in terms of the
two dimensional problem (for certain gauge field
backgrounds)\cite{FRS,FRS2}. Although its $T\to 0$ limit
coincides with the CS action, it has, at $T \ne 0$, a more
complicated form which guarantees gauge invariance at finite
temperature. It should be of interest to study these issues  in
the noncommutative case in view of possible applications of
noncommutative planar gauge theories to condensed matter
problems\cite{Sus,Fos}.

\section*{Acknowledgments}
I am grateful to the organizers of the  "II International Conference on
Fundamental Interactions'' and especially M.C.~Abdalla, J. Helayel Neto
and O.~Piguet (undoubtedly the heart of the meeting) for the splendid
days we spent in Pedra Azul. To all participants, thanks for the wonderful
and friendly atmosphere.
Partial support from UNLP, CICBA, ANPCyT (PICT 03-05179) is acknowledged.

\end{document}